\documentclass[5p]{elsarticle}

\usepackage{lineno,hyperref}
\modulolinenumbers[5]

\usepackage{graphics} % for pdf, bitmapped graphics files
\usepackage{epsfig} % for postscript graphics files
\usepackage{mathptmx} % assumes new font selection scheme installed
\usepackage{times} % assumes new font selection scheme installed
\usepackage{amsmath} % assumes amsmath package installed
\usepackage{amssymb}  % assumes amsmath package installed
\usepackage{caption}
\usepackage[figbotcap]{subfigure}
\usepackage{color}
\usepackage{gensymb}%for the \degree command

\usepackage{lipsum}
\makeatletter
\def\ps@pprintTitle{%
	\let\@oddhead\@empty
	\let\@evenhead\@empty
	\def\@oddfoot{}%
	\let\@evenfoot\@oddfoot}
\makeatother

\def\rd#1{{\color{red}{#1}}}
\def\pb#1{\footnote{PB:\rd{#1}}}
\def\eqdef{\ensuremath{:=}}

\def\mpcwh{\textit{SL-MPC}}
\def\mpcwoh{\textit{S-MPC}}
\def\base{\textit{BL}}

\usepackage{xparse}

\ExplSyntaxOn
\cs_new:Nn \pb_prop_gset_bykeys:Nn
{
	\prop_clear:N \l__pb_temp_prop
	\keys_set:nn { pb/propbykey } { #2 }
	\prop_gset_eq:NN #1 \l__pb_temp_prop
}
\cs_generate_variant:Nn \pb_prop_gset_bykeys:Nn { c }

\keys_define:nn { pb/propbykey }
{
	unknown .code:n = \prop_put:Nxn \l__pb_temp_prop { \l_keys_key_tl } { #1 }
}

\cs_generate_variant:Nn \prop_put:Nnn { Nx }

\NewDocumentCommand{\setupcollaborator}{mm}
{% #1 = identifier string, #2 = set of key-value pairs
	\prop_new:c { g_collaborator_#1_prop }
	\pb_prop_gset_bykeys:cn { g_collaborator_#1_prop } { #2 }
}

\NewDocumentCommand{\selectcollaborator}{m}
{
	\prop_map_inline:cn { g_collaborator_#1_prop }
	{
		\tl_set:cn { ##1 } { ##2 }
	}
}
\ExplSyntaxOff

\setupcollaborator{PBmac}
{% pb, macbook air or mac mini at work, 
	NAME=Prabir Mac , laptop or mac mini,
	DiCEbibPATH=/Users/pbarooah/Dropbox/DiCElab_bib,
	NRbibPATH = /Users/pbarooah/Dropbox/NSRbib,
}
\setupcollaborator{NRlinux}
{% NRlinux is Naren using the linux machine at work
	NAME=NR at lab,
	DiCEbibPATH=../../DiCElab_bib,
	NRbibPATH = ../../NSRbib,
}
\setupcollaborator{NRwindows}
{% NRwindows is Naren using his windows.
	NAME=NR using his windows computer,
	DiCEbibPATH=../../DiCElab_bib,
	NRbibPATH = ../../NSRbib,
}
\selectcollaborator{NRlinux} 
\graphicspath{{figures/}}

\begin{document}

\begin{frontmatter}

\title{MPC for Energy Efficient HVAC Control with Humidity and Latent Heat Considerations}

%\author{Naren Srivaths Raman}
%\author[mymainaddress]{Naren Srivaths Raman\corref{mycorrespondingauthor}}
%\ead{narensraman@ufl.edu}
%
%\author[mymainaddress]{Prabir Barooah}
%\ead{pbarooah@ufl.edu}
%
%\address[mymainaddress]{University of Florida, Gainesville, Florida USA}
%%\address[mysecondaryaddress]{360 Park Avenue South, New York}

\author{Naren Srivaths Raman\corref{mycorrespondingauthor}}
\cortext[mycorrespondingauthor]{Corresponding author}
\ead{narensraman@ufl.edu}
\author{Karthikeya Devaprasad\corref{}}
\ead{kdevaprasad@ufl.edu}
\author{Bo Chen\corref{}}
\ead{bo.chen@ufl.edu}
\author{Herbert A. Ingley\corref{}}
\ead{ingley@ufl.edu}
\author{Prabir Barooah\corref{}}
\ead{pbarooah@ufl.edu}
\address{Department of Mechanical and Aerospace Engineering, University of Florida, Gainesville, FL 32611, USA}
%	\author[mainaddress]{Naren Srivaths Raman\corref{mycorrespondingauthor}}

\begin{abstract}
Even though energy efficient climate control of buildings using model predictive control (MPC) has been widely investigated, most MPC formulations ignore humidity and latent heat. The inclusion of moisture makes the problem considerably more challenging, primarily since a cooling and dehumidifying coil model which accounts for both sensible and latent heat transfers is needed. In this work, we propose an MPC controller in which humidity and latent heat are incorporated in a principled manner. We construct low order data-driven models of a cooling and dehumidifying coil that can be used in the MPC formulation. The resulting controller's performance is tested in simulation using a plant that differs significantly from the model used by the optimizer. Additionally, the performance of the proposed controller is compared with that of a naive MPC controller which does not explicitly consider humidity, and also to that of a conventional rule-based controller. Simulations show that the proposed MPC controller outperforms the other two consistently. It is also observed that the naive MPC formulation which does not consider humidity leads to poor humidity control under certain conditions. Such violations in humidity can adversely affect occupant comfort and health.
\end{abstract}


\end{frontmatter}

%\linenumbers

\section{Introduction}\label{sec:intro}
The application of Model Predictive Control (MPC) for energy efficient climate control of buildings has been an active area of research; see the review articles \cite{serale2018model,shaikh2014review} and references therein. In MPC, control commands for a planning horizon are decided at every decision instant by solving an optimization problem, implementing only the first segment of the plan, and then repeating the process ad infinitum. Because of its use of numerical optimization, MPC can handle various constraints that are otherwise challenging to ensure, which has led to the success of MPC in many applications~\cite{qinbad:2003}.

In case of building climate control, the advantage of MPC over traditional rule-based controllers is that MPC can satisfy conflicting goals such as keeping energy use small while maintaining thermal comfort. Thermal comfort is influenced by several factors such as space temperature, humidity, air speed, clothing, metabolic rate, etc.~\cite{ashrae:2017thermal}. Space temperature and humidity are especially important factors in determining comfort and health~\cite{DudaPitfalls:2018,baughman1996indoor,fischer2003humidity}.  

Despite the importance of humidity and latent heat in building climate control, it is ignored in most existing MPC formulations. The principal challenge in including humidity and latent heat is that variables that determine the building's temperature and humidity---humidity and temperature of the conditioned air ---are a complex function of control commands, and cannot be independently chosen. The control commands that can be independently chosen are inlet conditions of the cooling coil that cools and dehumidifies the air supplied to the indoor space. Incorporating humidity into MPC requires a model of the cooling and dehumidifying coil that accounts for both \emph{sensible and latent heat transfers}, and predicts how control commands (conditions at the coil inlet) determines the temperature and humidity of the conditioned air. Such models are usually highly complex. Some are partial differential equations (PDEs) with a large number of parameters and several sub-models  based on the condition of the cooling coil such as completely dry, completely wet, and partly wet~\cite{zhou2007simplified}. Some are ordinary differential equations or even static models consisting of a large number of empirical relations that vary depending on coil geometry, configuration, and manufacturer~\cite{doe:2018energyplus,klein2009trnsys}.  Such complex models are not suitable for MPC, which involves real-time optimization. In addition, nonlinearities in the humidity dynamics make the underlying optimization problem non-convex~\cite{SG_PB_Energy:11}.

Rule-based controllers that are currently used in practice employ conservatively designed rules that have been arrived at after decades of experience. For instance, a widely used heuristic in hot humid climates is to keep the conditioned air setpoint at 12.8~\degree C (55~$\degree$F)~\cite{williams2013why}. This low value ensures the air delivered to the indoor space is dry enough to maintain humidity within allowable limits under worst-case conditions. The downside is that worst-case conditions occur rarely, which leads to high energy use. Not only is the air cooled unnecessarily, but it must then be reheated to prevent the indoor space from becoming too cold.

The recent literature on MPC for HVAC control is focused on energy use minimization while maintaining thermal comfort and indoor air quality~\cite{serale2018model,shaikh2014review}. The motivation is the large energy footprint of HVAC systems. An MPC controller which minimizes energy/cost without including humidity and latent heat in its problem formulation could have two potential issues. One, it may lead to poor humidity control. Two, since the latent component of cooling---energy required to dehumidify air---is not accounted for in the objective function, the predicted energy use by the controller may be far from the actual energy use when the controller is used in practice. 
%Three, one of the easiest ways to reduce the humidity of air is by cooling it, which is how most HVAC systems in buildings operate. Therefore, there could be situations where simultaneous cooling and heating is inevitable to meet the ventilation requirements and comfort limit constraints. Such behavior is not possible by design in an MPC controller that does not include humidity.

In this paper, we propose an MPC formulation for energy efficient climate control of a building in which humidity and latent heat are taken into account in a principled manner. The proposed controller is hereafter referred to as $\mpcwh$, because it accounts for both sensible and latent components of cooling. We specifically focus on a variable-air-volume (VAV) heating, ventilation and air conditioning (HVAC) system that uses chilled water to cool and dehumidify, i.e., condition the air supplied to the building. Figure \ref{fig:AHU_chiller_schematic} shows the schematic of a a VAV HVAC system. To avoid clutter, we consider a single zone building, though the proposed method can be  extended to multi-zone buildings.
 
As mentioned earlier, one of the main challenges of including humidity and latent heat is the need for a cooling and dehumidifying coil model. To address this challenge we develop a data-driven low order model that predicts temperature and humidity of the conditioned air (outputs) as a function of the inputs: the temperature, humidity and flow rate of air incident on the coil, and temperature and flow rate of chilled water entering the coil. This model is used in the optimizer used by the MPC controller. We also develop a slightly more complex, but much higher accuracy, data-driven model that is used to simulate the plant. Both models are identified from data, which can come from experiments or from software such as EnergyPlus~\cite{crawley2001energyplus}. 

Apart from the proposed MPC controller and the data-driven cooling coil models, a third contribution of the paper is a comparison of the performance of the proposed controller with two other controllers: (i) an MPC controller that does not have humidity constraints and latent heat explicitly accounted for, which is referred to as $\mpcwoh$ since it only accounts for sensible heat, and (ii) a widely used rule-based controller (``single maximum'' \cite{ASHRAE_handbook_applications:11}), which is referred to as $\base$ (for \emph{baseline}).

The simulation studies reported here show that the proposed $\mpcwh$ controller uses the least amount of energy and meets thermal comfort constraints as well or better, compared to the other two controllers. It is also observed that $\mpcwoh$ makes decisions that either lead to poor humidity control under certain conditions. For instance, on a spring or summer night, it decides that slightly cooler outside air can provide free cooling, but the high humidity of the outside air makes indoor humidity higher than what should be allowed. Such ``optimal'' decisions over long periods of time can lead to issues such as mold growth, a critical health concern in hot humid climates~\cite{baughman1996indoor,fischer2003humidity}. Therefore, ignoring humidity in MPC problem formulations, especially for such conditions, should be avoided.

A preliminary version of this work has been reported in \cite{RamanMPCACC:2019}, wherein we have compared the performance of $\mpcwh$ and $\base$. In this paper, we add $\mpcwoh$ to the comparison and quantify their performance in terms of energy consumption and thermal comfort violation. This comparison clarifies the implication of ignoring humidity in the problem formulation: that serious adverse effects can occur. 

%There are a few challenges associated with implementing $\mpcwh$: (i) input output data is needed to fit the parameters in the cooling and dehumidifying coil model, (ii) computational complexity increases as there are more decision variables in the underlying optimization problem being solved, (iii) humidity sensors need to be installed in the zones and at certain locations in the air loop.

 %Another important situation when $\mpcwoh$ fails to make the right decisions is under high internal heat load and mild outdoor conditions. Even though comfort limit violation is not an issue, the lack of latent heat in the MPC problem formulation causes the controller to pick an action that is energy inefficient, when compared to $\mpcwh$. Interestingly, $\mpcwoh$ makes decisions similar to $\mpcwh$ under certain other conditions. At high internal heat load and hot outdoor weather conditions, the controller recognizes that the conditioned air temperature must be low enough to maintain the indoor temperature within allowable limits. That decision has an unintended but good side effect of maintaining space humidity. Finally, under cold outdoor weather conditions, humidity is not a concern since the outdoor air is dry, and it can be used for ``free cooling''.

The rest of this paper is organized as follows. Section~\ref{sec:review} reviews the related work on MPC-based building climate control vis-\`a-vis humidity and latent heat considerations. Section~\ref{sec:system_models} describes a single-zone variable air volume (VAV) HVAC system and the mathematical models we use in simulating the plant (the system to be controlled). Section~\ref{sec:control_algorithms} presents the  proposed $\mpcwh$ control strategy, and the control-oriented cooling and dehumidifying coil model. It also describes the two other control algorithms used for comparison with the proposed controller. The simulation setup and their results are discussed in Sections~\ref{sec:sim_setup} and \ref{sec:results_discussions} respectively. Concluding remarks are provided in Section~\ref{sec:conclusion}.

\subsection{Review of prior work} \label{sec:review}

There have been several studies in which MPC is used for energy efficient climate control of buildings---see the review papers~\cite{serale2018model,shaikh2014review} and the references therein.  However, there are only a few works which have considered humidity explicitly in their MPC problem formulation. We limit ourselves to these references. Based on the objective function to be minimized in the MPC formulation, these works can be classified into two broad categories: (i) economic MPC and (ii) set point tracking MPC. In set point tracking MPC, the objective function is chosen so that minimization of the objective function helps to drive the relevant output(s) to the desired set point. In economic MPC, the objective function is chosen to be a performance measure---usually the economic cost---that may not correspond to a steady state operation as it does in case of set point tracking. See~\cite{rawlings2017model,rawlings2012fundamentals} for a through exposition of tracking and economic MPC.

References~\cite{schwingshackl2016LoLiMoT,xi2007support,wang2013desiccant} are examples of setpoint tracking MPC. 
In \cite{schwingshackl2016LoLiMoT}, an MPC controller is designed to maintain the supply air temperature and humidity at a given set point by varying the mass flow rate of chilled water and inlet water temperature of the heating coil. In \cite{xi2007support}, an aggregated model of the building and HVAC system is obtained with the supply air fan speed and the chilled water valve opening as inputs, and room temperature and relative humidity as outputs. Subsequently, an MPC controller is used to maintain the room temperature and relative humidity at its set point with the above model. Both temperature and humidity are considered in the problem formulation. A control-oriented desiccant wheel model is used in an MPC based control scheme to regulate humidity in \cite{wang2013desiccant}.

The MPC controller proposed in this paper, and those in references~\cite{goybar:CDC:2013,SG_HI_PB_AE:2013,kumar2010design} belong to the category of economic MPC, with total energy use being the objective function to minimize. In \cite{goybar:CDC:2013}, it is assumed that the relative humidity of the conditioned air after the cooling coil is always 90\%, while \cite{SG_HI_PB_AE:2013} assumes both the temperature and the humidity ratio of the conditioned air are constant. These assumptions avoid the need for a cooling coil model though the validity of these assumptions is questionable. An economic MPC scheme---for energy use minimization---with humidity and latent heat considerations is presented in~\cite{kumar2010design}. Unlike the chilled water system used in this work, the focus in \cite{kumar2010design} is on direct expansion (DX) cooling systems.

There are also a few papers in which the terms in the objective function consist of both energy use and deviation from set points, so these can be thought of as a hybrid between tracking and economic MPC---\cite{zavala2012real,wang2000model}. Multiple MPC strategies are compared for an air handling unit serving a single-zone in~\cite{zavala2012real}. It is assumed that the temperature and humidity ratio after the cooling coil can be chosen independently, thereby not requiring the use of a cooling coil model. This assumption will not hold in physical systems, as the only variables that can be independently chosen are the inlet conditions to the coil.

Ref.~\cite{wang2000model} is the most relevant to our work; they use a cooling coil model in their optimization in which temperature and humidity of the conditioned air is modeled correctly to be thermodynamically coupled. The supply air flow rate is not a control command, while in our formulation it is. The controller in~\cite{wang2000model} will be unaware of disturbances in the longer time scales, since a short prediction horizon of 10 minutes is used. In contrast, we use a prediction horizon of 24 hours. Moreover, there are multiple elements included in the objective function: energy use, thermal comfort, indoor air quality, etc., which needs careful tuning of weights. In our formulation, energy use is the objective to be minimized, with thermal comfort and indoor air quality being constraints to be met. Lastly, a nondeterministic optimization algorithm (genetic algorithm) is used to perform the minimization while we use a deterministic search method through a nonlinear programming (NLP) solver.

Although the papers on HVAC control that do not consider humidity and latent heat are outside the scope of this review, a subset of those works report experimental evaluations in real buildings. These deserve special attention: if an MPC controller that does not consider humidity and latent heat can still provide good performance in real buildings that are affected by humidity and latent heat, then incorporating these features into the controller---which necessarily increase complexity---is perhaps not necessary. In particular, refs.~\cite{sturzenegger2016model,Siroky:AE:2011,Bengeaetal:2013} describe experimental demonstrations that have been carried out with MPC-based controllers on real buildings. The problem formulations in these references do not consider latent heat/room humidity dynamics. It is not clear from the reported assessment if the controllers were able to maintain humidity, since humidity measurements were not reported. 

In \cite{sturzenegger2016model} an MPC based controller was implemented in a Swiss office building. They used thermally activated building systems, an air handling unit, and blinds, for actuation.  Majority of the experiments were done when the weather was cold and dry in which humidity and latent cooling loads were unlikely to be of concern. However, one set of experiments was done between May-August when it was hot and humid. Space humidity was not reported in the evaluations. %Since the climatic condition in Switzerland is dry, and \rd{especially so during the time of the experiments??}, it is perhaps reasonable to ignore the effects of humidity. \rd{14 weeks of cooling season (May to Aug 2012), 6 weeks of heating season (Nov-Dec 2012), 9 weeks of heating season (Dec 2012-Mar 2013).} 
The MPC demonstration reported in \cite{Siroky:AE:2011} controlled the heating system of a building in Prague during winter when humidity is not a concern for that climate.

The work \cite{Bengeaetal:2013} describes an MPC-based controller that was implemented in a mid-size ($650 m^2$) commercial building in Champaign, Illinois, which is hot and humid during the summer. Two sets of tests were conducted. One was during the transition season in October (2011) and the other was during cold season (February 2012). It is not clear from published results whether humidity was maintained within acceptable limits, since only zone temperature and $CO_2$ levels were reported, not humidity.

In summary, it is not possible to say from the published literature if an MPC controller that does not consider humidity and latent heat is able to provide humidity control. Our results---reported later in the paper---indicate it is unlikely in hot and humid climates, thus motivating a need for an MPC formulation that includes these features. 

%Ref.~\cite{rawlings2018economic} reports work that goes beyond demonstration; it is a commercial implementation of MPC to building control. However, the focus of that work is optimal scheduling of the central plant in a university campus;  climate control of individual buildings is performed by a traditional system that maintains humidity through conservative rule-based control. 

\section{System description and models}\label{sec:system_models}

\begin{figure}[t]
	\centering
	\includegraphics[width=0.99\linewidth]{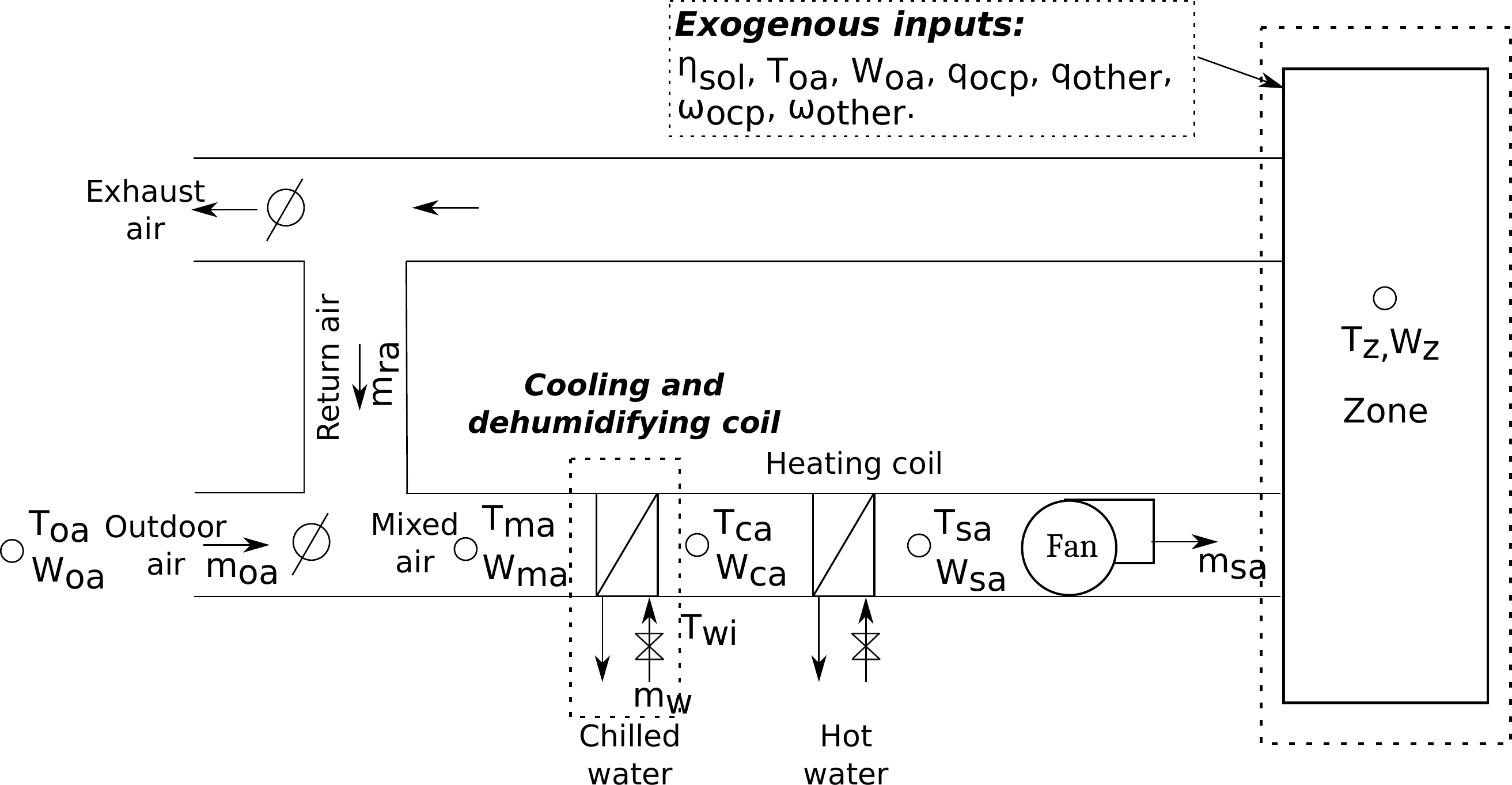}
	\caption{Schematic of a single-zone commercial variable-air-volume
		HVAC system.}
	\label{fig:AHU_chiller_schematic}
\end{figure}

Figure~\ref{fig:AHU_chiller_schematic} shows the schematic of a typical single-zone variable-air-volume HVAC system commonly used in commercial buildings. In such a system, part of the air exhausted from the zone is recirculated and mixed with outdoor air. Then the mixed air is sent through a cooling coil where it is cooled and dehumidified to conditioned air temperature ($T_{ca}$) and humidity ratio ($W_{ca}$). This air is then passed through a reheat coil where the air is heated to supply air temperature ($T_{sa}$) before being supplied to the zone. There is no water vapor phase change across the heating coil, so the humidity ratio of supply air and conditioned air is the same: $W_{sa}=W_{ca}$. \emph{The role of the climate control system is to vary the following control commands: (i) supply air flow rate ($m_{sa}$), (ii) outdoor air ratio ($r_{oa}$, which is the ratio of outdoor air flow rate to supply air flow rate, $r_{oa}=\frac{m_{oa}}{m_{sa}}=\frac{m_{oa}}{m_{oa}+m_{ra}}$), (iii) conditioned air temperature ($T_{ca}$), and (iv) supply air temperature ($T_{sa}$), to maintain thermal comfort and indoor air quality in the zone. So the control command vector is:}\vspace{-15pt}

\begin{align}\label{eq:control_command}
	 u=[m_{sa},r_{oa},T_{ca},T_{sa}]^T\in R^4.
 \end{align}

In the following subsections we describe the mathematical models that are used to simulate the plant (the dynamic system being controlled). The plant configuration and parameters are chosen to mimic a real HVAC system, the one that serves a $465$~$m^2$ (5000 sq.ft.) auditorium in Pugh Hall at the University of Florida campus. 

\subsection{Hygro-thermal dynamics model}\label{section:hygrothermal}
We use the following RC (resistor-capacitor) network model for the temperature dynamics of the zone serviced by the HVAC system~\cite{cofbar:2018}:
\begin{align}\label{eq:thermal_dynamic}
C_z\dot{T}_z(t) &=  \frac{(T_{w}(t) - T_z(t))}{R_w} + q_{HVAC}(t)+A_e\eta_{sol}(t) + \nonumber \\
& \quad q_{ocp}(t) + q_{other}(t) \\
C_w\dot{T}_w(t) &= \frac{(T_{oa}(t) - T_w(t))}{R_z} + \frac{(T_z(t) - T_w(t))}{R_w} 
\end{align}
where $T_z$ is the zone temperature, $T_w$ is the wall temperature, $T_{oa}$ is the outdoor air temperature, $q_{HVAC}$ is the heat influx due to the HVAC system, $\eta_{sol}$ is the solar irradiance, $q_{ocp}$ is the occupant-induced internal heat load, $q_{other}$ is the heat load due to other sources such as lighting and equipment, $C_z$ and $C_w$ are the thermal capacitance of the zone and the wall respectively, $R_z$ is the resistance to heat exchange between the outdoors and wall,  $R_w$ is the resistance to heat exchange between the wall and indoors and $A_e$ is the effective area of the building. The heat influx due to the HVAC system is a function of the supply air temperature and zone temperature:
\begin{align}\label{eq:q_HVAC}
q_{HVAC}(t) = m_{sa}(t)C_{pa}(T_{sa}(t)-T_z(t)) ,
\end{align}
where $m_{sa}$ is the supply air flow rate and $C_{pa}$ is the specific heat of air at constant pressure.

The dynamics of zone humidity ratio $W_z$ is modeled as: 
\begin{align}\label{eq:humidity_dynamic}
\dot{W}_z(t) = \frac{R_gT_z(t)}{VP^{da}}\Bigg[\omega_{ocp}(t) + \omega_{other}(t) + m_{sa}(t)\frac{W_{sa}(t)-W_z(t)}{1+W_{sa}(t)}\Bigg]
\end{align}
where $V$ is the zone volume, $R_g$ is the specific gas constant of dry air, $P^{da}$ is the partial pressure of dry air, $W_{sa}$ is the supply air humidity ratio, $\omega_{ocp}$ and $\omega_{other}$ are the rate of internal water vapor generation due to people and other sources,  respectively~\cite{SG_PB_Energy:11}.

\subsection{Cooling and dehumidifying coil model}\label{section:cooling_coil}
\begin{figure}[t]
	\centering
	\includegraphics[width=0.44\linewidth]{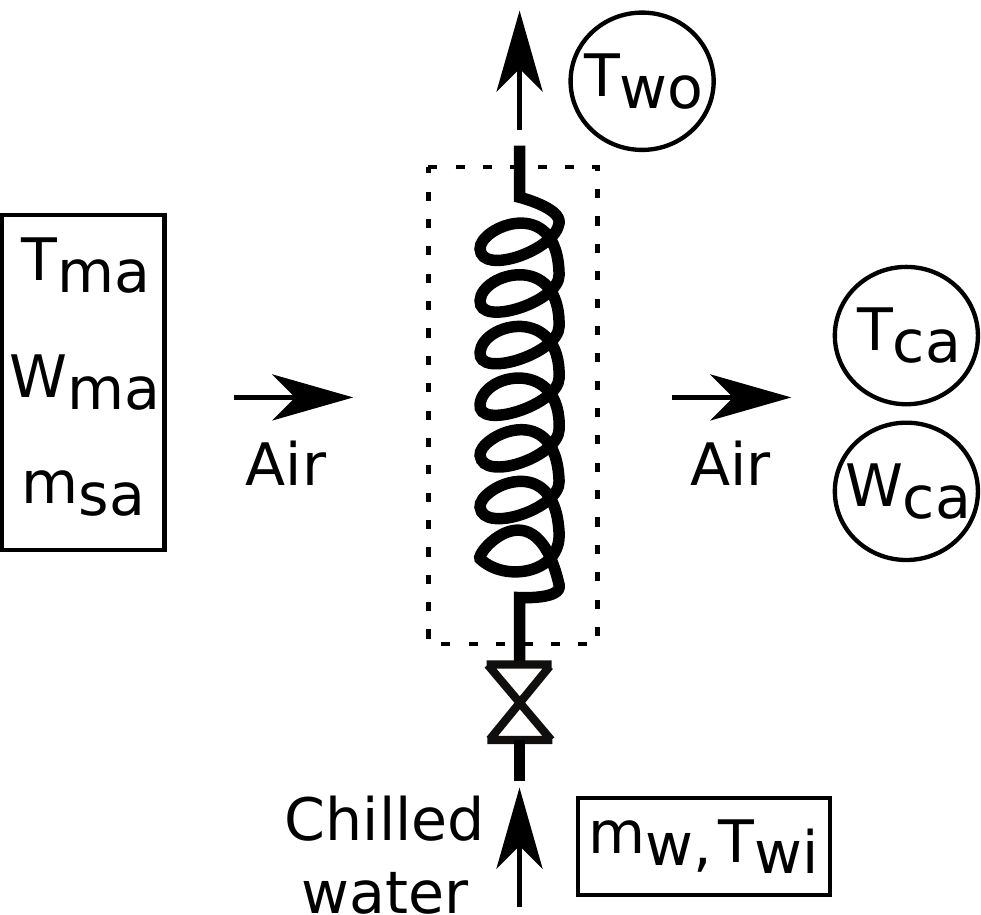}
	\caption{A cooling and dehumidifying coil, and relevant variables (model inputs in rectangles, outputs in circles).}
	\label{fig:cooling_coil}
\end{figure}

The inputs for the model are supply air flow rate ($m_{sa}$), mixed air temperature ($T_{ma}$), mixed air humidity ratio ($W_{ma}$), chilled water flow rate ($m_w$), and inlet water temperature ($T_{wi}$); see Figure~\ref{fig:cooling_coil}. The outputs are conditioned air temperature ($T_{ca}$) and humidity ratio ($W_{ca}$). %Usually the inlet water temperature is maintained at a constant value and therefore the outputs of our model are a function of the remaining four inputs.

There is a rich literature on modeling cooling and dehumidifying coils; see \cite{zhou2007simplified,klein2009trnsys} and references therein. However, some of these models require coil geometry data which is hard to obtain. Another class of models involve complex partial differential equations~\cite{zhou2007simplified}. For our purposes a simple static model would suffice as the time constants for a cooling coil are small---about 60 to 120 seconds (see Figures~4 to 7 in  \cite{zhou2007simplified})---compared to the time constant of zone thermal dynamics, which is in hours~\cite{cofbar:2018}. The model used in EnergyPlus (see Section 16.2.1 in \cite{doe:2018energyplus}) is such a static model. It is still complex and difficult to replicate as it involves many empirical relations. Therefore, we opt for a grey box data-driven model. EnergyPlus is used as a ``virtual cooling coil testbed", and data collected from EnergyPlus simulations is used to fit the parameters of the model. The process is explained below.

A single-zone commercial building is simulated in EnergyPlus version 8.9 \cite{crawley2001energyplus}, with a cooling coil pulling in unmixed outdoor air and supplying it to the zone after cooling and dehumidifying it. Using unmixed air ensures that we have full control over the temperature and humidity ratio of air entering the cooling coil, as EnergyPlus allows the use of a custom generated weather file to specify outdoor conditions. The HVAC air loop also contains a variable flow fan motor to control the mass flow rate of air, and the plant loop contains an electric chiller with variable flow pump to control the mass flow rate of water. The inlet and outlet conditions of the cooling coil are measured.

\begin{figure}[t!]
	\subfigure[][Measured ($T_{ca}$, output from EnergyPlus simulations) and predicted ($\hat{T}_{ca}$) value of conditioned air temperature for a specific bin, $T_{ma}=23.9~\degree C$ (75~\degree$F$) and $RH_{ma}=50\%$.\label{fig:T_ca_binned_75F_50per}]{\includegraphics[width=0.46\textwidth]{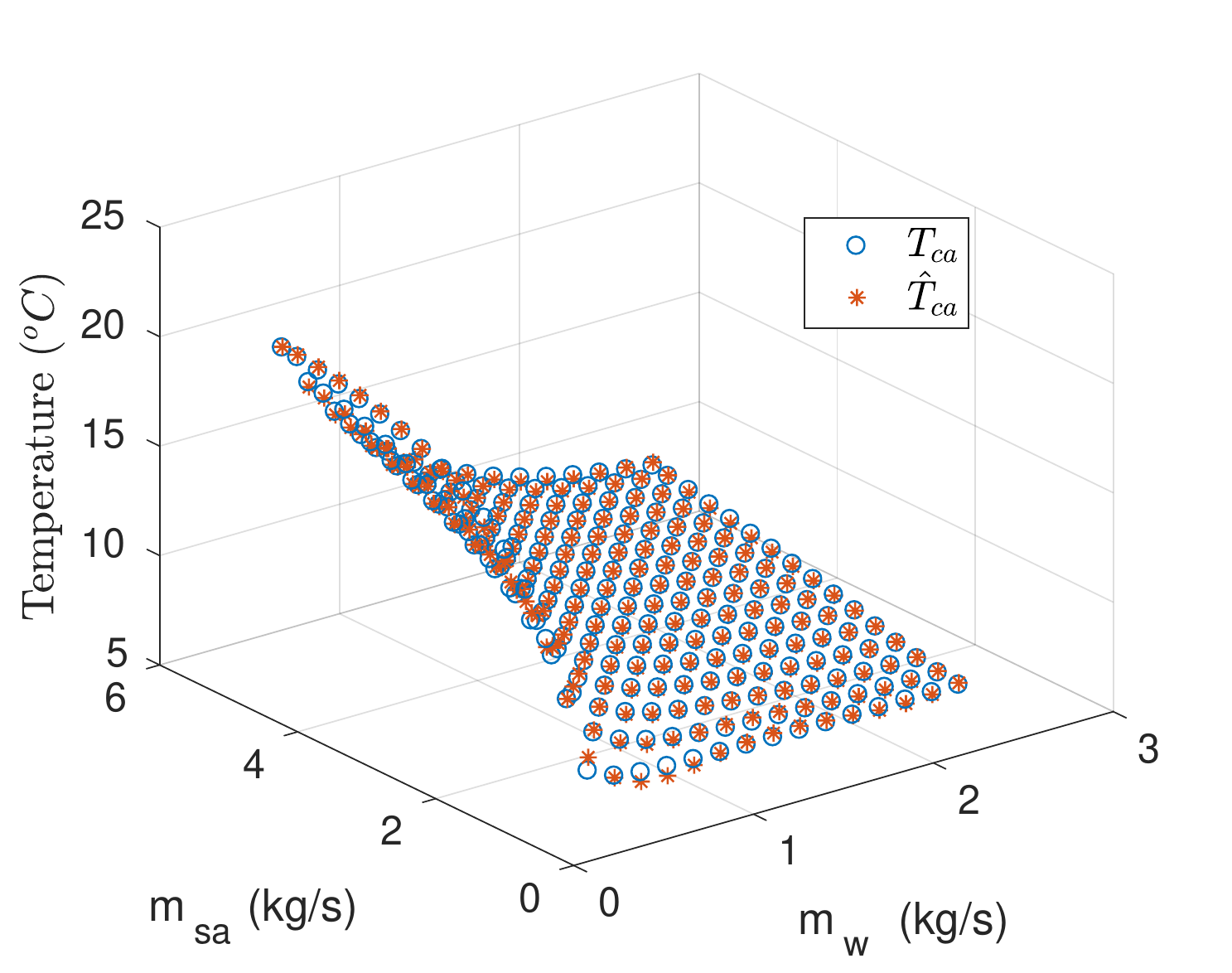}} \hfill
	\subfigure[junk][Measured ($W_{ca}$, output from EnergyPlus simulations) and predicted ($\hat{W}_{ca}$) value of conditioned air humidity ratio for a specfic bin, $T_{ma}=23.9~\degree C$ (75~\degree$F$) and $RH_{ma}=50\%$.\label{fig:W_ca_binned_75F_50per}]{\includegraphics[width=0.46\textwidth]{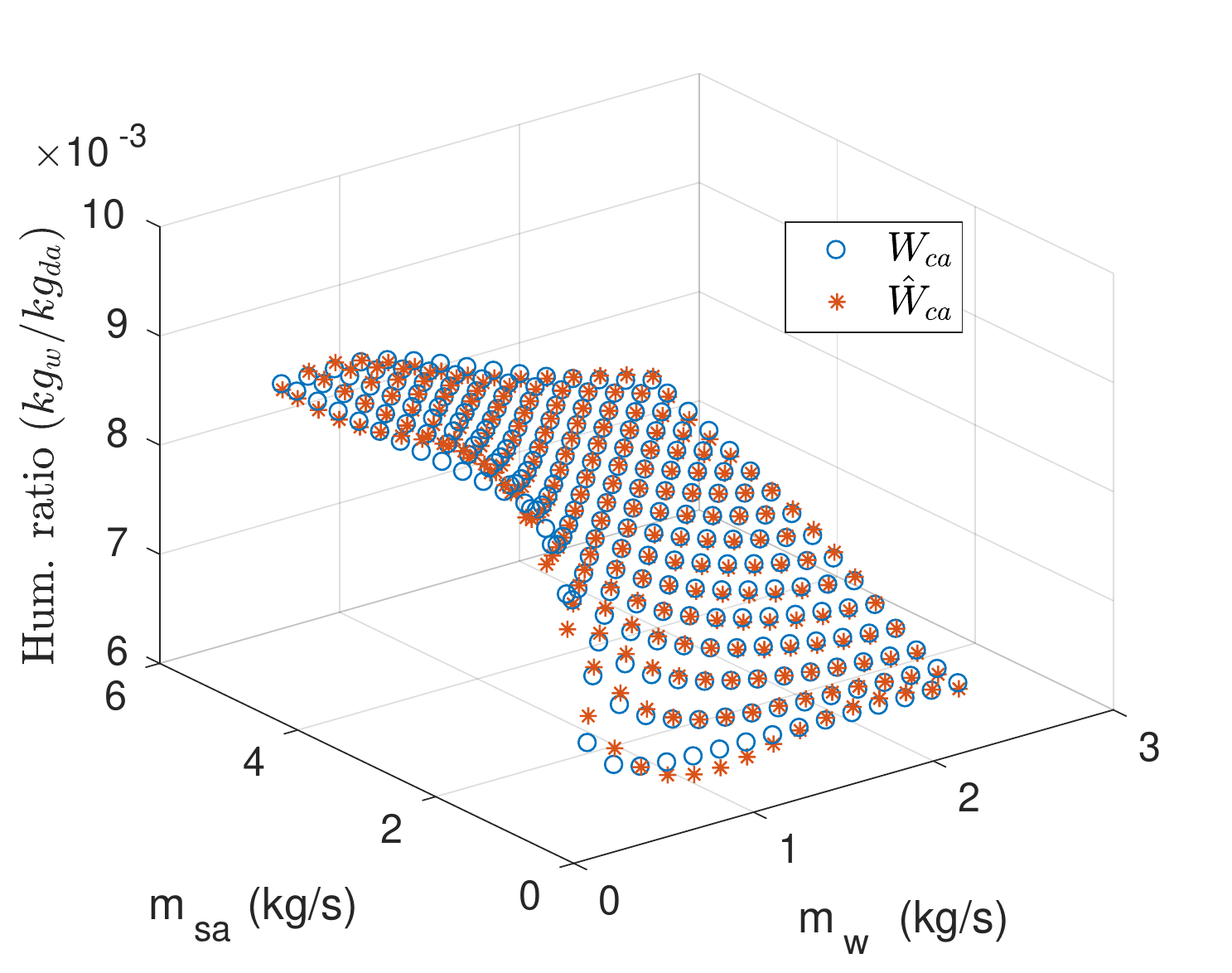}}
	\caption{Cooling coil binned model (used in simulating the plant).}
	\label{fig:cooling_coil_binned}
\end{figure}

The rates of flow through the pump and fan are varied using Building Controls Virtual Test Bed (BCVTB) \cite{wetter2008modular}. The air flow rate is varied from 0.1705~$kg/s$ (300 $ft^3/min$) to 4.6 $kg/s$ (8100 $ft^3/min$) and the water flow rate is varied from 0 $kg/s$ (0 $gallons/minute$) to 2.21 $kg/s$ (35 $gallons/minute$). The limits are chosen to mimic the equipment in Pugh Hall. The temperature and humidity ratio of outdoor air are controlled using a custom weather file. Since there are no other components before the coil that interact with outdoor air, we are able to use it to modulate the input conditions to the coil. The temperature is varied from 10 $\degree C$ (50~$\degree F$) to 43.3 $\degree C$ (110~$\degree F$) with steps of 0.56~$\degree C$ (1~$\degree F$). The relative humidity is varied from 10 \% to 100 \% with steps of 5 \%. The model is calibrated using 296,704 data points generated by varying inputs. A separate data set is generated for model validation.

We observe from our initial attempts that for a \emph{fixed} mixed air temperature and relative humidity, the outputs $T_{ca}$ and $W_{ca}$ can be predicted quite well by modeling them as polynomial functions of the mass flow rates of chilled water and supply air. Figure~\ref{fig:cooling_coil_binned} shows an example, using a 5$^{th}$ degree polynomial. However, a single polynomial leads to large errors when used at different mixed air temperatures and relative humidities. We therefore bin the inputs according to $T_{ma}$ and $RH_{ma}$ into 1159 bins, and use a 5$^{th}$ degree polynomial model for each bin. The resulting model is called a ``binned model''. The root mean square error for the validation data is less than $0.28~\degree C$ (0.5 \degree $F$, 1\%) for $T_{ca}$ and $0.3\times10^{-4}~kg_w/kg_{da}$ (1\%) for $W_{ca}$.

\subsection{Power consumption models}\label{section:power}
We assume that the power consumed by components such as dampers is negligible; the only power consuming components are the air supply fan, the reheat coil, and the cooling coil. The fan power is usually modeled as a quadratic function of the supply air flow rate~\cite{RamanAnalysisHPB:2018}:
\begin{align} \label{P_fan}
P_{fan} = \alpha_fm_{sa}(t)^2.
\end{align}
The power consumed by the cooling and dehumidifying coil is modeled as being proportional to the heat it extracts from the mixed air stream as follows:
\begin{align} \label{P_cc}
P_{cc}(t) = \frac{m_{sa}(t)\big[h_{ma}(t)-h_{ca}(t)\big]}{\eta_{cc}COP_c},
\end{align}
where $h_{ma}(t)$ and $h_{ca}(t)$ are the specific enthalpies of the mixed and supply air respectively, $\eta_{cc}$ is the cooling coil efficiency, and $COP_c$ is the chiller coefficient of performance. Since a part of the return air is mixed with the outside air, the specific enthalpy of the mixed air is:
\begin{align} \label{h_ma}
h_{ma}(t) &= r_{oa}(t)h_{oa}(t) + (1-r_{oa}(t))h_z(t),
\end{align}
where $h_{oa}(t)$ and $h_z(t)$ are the specific enthalpies of the outdoor and zone air respectively, and $r_{oa}(t)$ is the outside air ratio: $r_{oa}(t)\eqdef \frac{m_{oa}(t)}{m_{sa}(t)}$. The specific enthalpy of moist air with temperature $T$ and humidity ratio $W$ is given by \cite{ASHRAE_handbook_fund:17}: $	h(T,W) = C_{pa}T + W(g_{H_20}+C_{pw}T)$,
%\begin{align}\label{eq:enthalpy}
%	h(T,W) = C_{pa}T + W(g_{H_20}+C_{pw}T)
%\end{align}
where $g_{H_20}$ is the heat of evaporation of water at 0~$\degree C$, and $C_{pa},C_{pw}$ are specific heat of air and water at constant pressure.

The power consumed by the reheat coil is modeled as being proportional to the heat added to the conditioned air stream by the coil. Since the humidity ratio does not change across the reheat coil ($W_{sa}=W_{ca}$), the power consumption has the form
\begin{align} \label{P_reheat}
P_{reheat}(t) &= \frac{m_{sa}(t)C_{pa}\big[T_{sa}(t)-T_{ca}(t)\big]}{\eta_{reheat}COP_{h}}.
\end{align}
where $\eta_{reheat}$ is the reheat coil efficiency, and $COP_h$ is the boiler coefficient of performance.

\section{Control algorithms}\label{sec:control_algorithms}

\subsection{Proposed  controller: $\mpcwh$}\label{section:MPC}
\begin{figure}[t]
	\centering
	\includegraphics[width=0.98\linewidth]{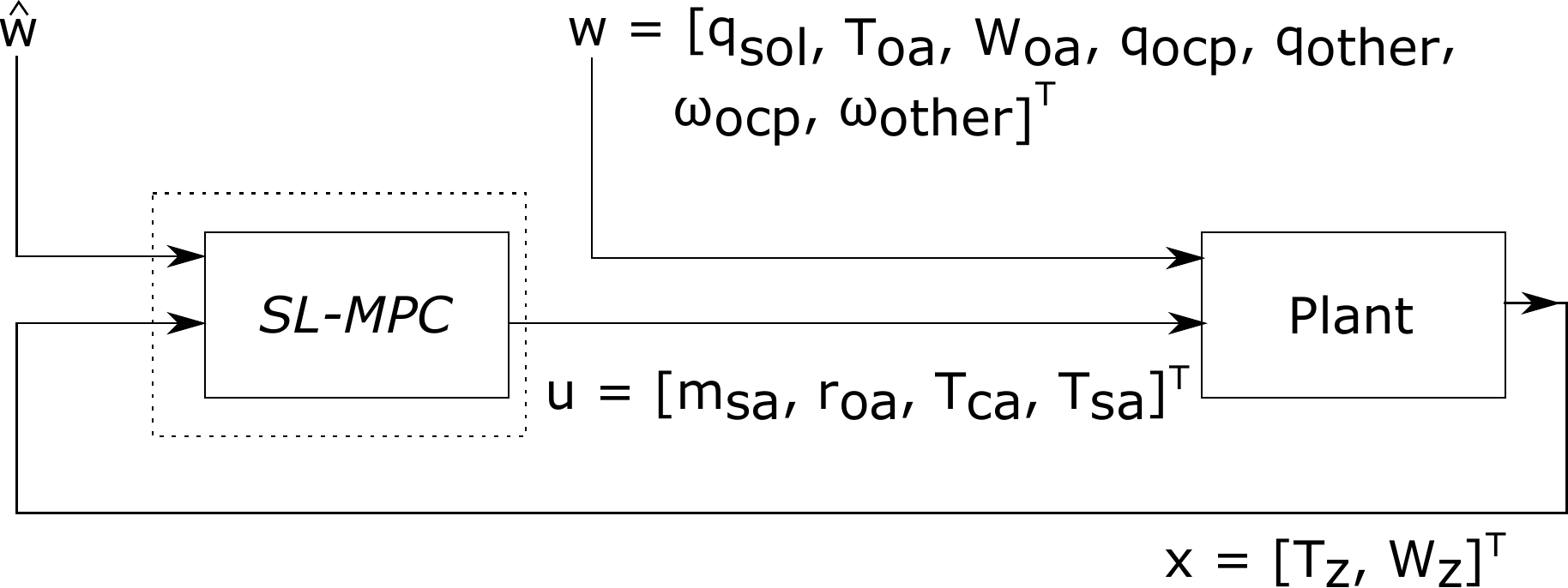}
	\caption{Proposed $\mpcwh$, control architecture.}
	\label{fig:MPC_Schematic}
\end{figure}

Figure~\ref{fig:MPC_Schematic} shows the control architecture for the proposed $\mpcwh$ controller. Control decisions are computed in discrete time indices $k=0,1,\dots$, with $\Delta t$ being the sampling interval. 

The control inputs for $N$ time steps are obtained by solving a constrained optimization problem of minimizing the energy consumption subject to thermal comfort, indoor air quality, and actuator constraints. Then the control inputs obtained for the first time step are applied to the plant. The optimization problem is solved again for the next $N$ time steps with the initial state of the model obtained from plant measurements. This process is repeated at the next time instant. To describe the optimization problem, first we define the state vector $x(k)$ and the vector of control commands and internal variables $v(k)$ as: 
\begin{align*}
&x(k) := [T_z(k), W_z(k)]^T, \\
&v(k) := [u(k)^T, m_w(k), W_{ca}(k)]^T, 
\end{align*}
where $u(k)$ is the control command vector defined in \eqref{eq:control_command}. The exogenous input vector is:
\begin{align*}
&w(k) := [\eta_{sol}(k), T_{oa}(k), W_{oa}(k), q_{ocp}(k), q_{other}(k), \omega_{ocp}(k), \\&\qquad\omega_{other}(k)]^T.
\end{align*}
At time index $j$, the decision variables in the optimization problem underlying the proposed MPC controller are denoted by $X$ and $V$, where $X=[x^T(j+1),x^T(j+2),\dots,x^T(j+N)]^T$ and $V=[v^T(j),v^T(j+1),\dots,v^T(j+N-1)]^T$. The predictions of the exogenous inputs $W = [w^T(j),w^T(j+1),\dots,w^T(j+N-1)]^T$ are assumed known at time index $j$. In simulations reported later, we use $\Delta t=5$ minutes and prediction/planning horizon of $N=288$ (corresponding to 24 hours).

The optimization problem at time index $j$ is:
\begin{align} \label{eq:cost}
\min_{V,X} \sum\limits^{j+N-1}_{k=j} \bigg[P_{fan}(k) + P_{cc}(k) + P_{reheat}(k)\bigg] \Delta t,
\end{align}
where $P_{fan}$, $P_{cc}$ and $P_{reheat}$ are given by \eqref{P_fan}, \eqref{P_cc} and \eqref{P_reheat} respectively, and is subject to the following constraints:
\begin{align}
&T_z(k+1)=T_z(k) + \frac{\Delta t}{C}\bigg[\frac{(T_{oa}(k) - T_z(k))}{R} + q_{HVAC}(k) + \nonumber\\ &\qquad \qquad \quad A_e\eta_{sol}(k)+q_{ocp}(k)+q_{other}(k)\bigg], \label{eq:T_z_eq}
\end{align}
\begin{align}
&W_z(k+1) = W_z(k)+\frac{\Delta t R_gT_z(k)}{VP^{da}}\bigg[\omega_{ocp}(k) + \omega_{other}(k) + \nonumber\\ &\qquad \qquad \quad m_{sa}(k)\frac{W_{sa}(k)-W_z(k)}{1+W_{sa}(k)}\bigg], \label{eq:W_z_eq} \\
&T_{ca}(k) = f_{co}\big(T_{ma}(k),W_{ma}(k),m_{sa}(k),m_w(k)\big), \label{eq:cc_T_ca}\\
&W_{ca}(k) = g_{co}\big(T_{ma}(k),W_{ma}(k),m_{sa}(k),m_w(k)\big), \label{eq:cc_W_ca}\\
&T_z^{low}(k) \leq T_z(k) \leq T_z^{high}(k), \label{eq:T_z_ineq}\\
&W_z^{low}(k) \leq W_z(k) \leq W_z^{high}(k), \label{eq:W_z_ineq}\\
&m_{sa}(k+1) \leq min\big(m_{sa}(k)+m_{sa}^{rate}\Delta t,m_{sa}^{high}\big), \label{eq:m_sa_ineq1}\\
&m_{sa}(k+1) \geq max\big(m_{sa}(k)-m_{sa}^{rate}\Delta t,m_{sa}^{low}\big), \label{eq:m_sa_ineq2}\\
&r_{oa}(k+1) \leq min\big(r_{oa}(k)+r_{oa}^{rate}\Delta t,r_{oa}^{high}\big), \label{eq:r_oa_ineq1}\\
&r_{oa}(k+1) \geq max\big(r_{oa}(k)-r_{oa}^{rate}\Delta t,r_{oa}^{low}\big), \label{eq:r_oa_ineq2}\\
&T_{ca}(k+1) \leq min\big(T_{ca}(k)+T_{ca}^{rate}\Delta t,T_{ma}(k+1)\big), \label{eq:T_ca_ineq1}\\
&T_{ca}(k+1) \geq max\big(T_{ca}(k)-T_{ca}^{rate}\Delta t,T_{ca}^{low}\big), \label{eq:T_ca_ineq2}\\
&T_{sa}(k+1) \leq min\big(T_{sa}(k)+T_{sa}^{rate}\Delta t,T_{sa}^{high}\big), \label{eq:T_sa_ineq1}\\
&T_{sa}(k+1) \geq max\big(T_{sa}(k)-T_{sa}^{rate}\Delta t,T_{ca}(k+1)\big), \label{eq:T_sa_ineq2}\\
&W_{ca}(k)\leq W_{ma}(k), \label{eq:W_ca_ineq}
\end{align}
where constraints \eqref{eq:T_z_eq}-\eqref{eq:cc_W_ca} and \eqref{eq:W_ca_ineq} are for $k=j,...,j+N-1$, constraints \eqref{eq:T_z_ineq} and \eqref{eq:W_z_ineq} are for $k=j+1,...,j+N$, and constraints \eqref{eq:m_sa_ineq1}-\eqref{eq:T_sa_ineq2} are for $k=j,...,j+N-2$.
%We can see that \eqref{P_cc} uses the enthalpies of the mixed and conditioned air to calculate the power consumed by the cooling coil. Therefore, it accounts for both the \emph{latent and sensible} thermal power components.

The constraint \eqref{eq:T_z_eq} is due to the thermal dynamics of the zone, which is a discretized form of a first-order RC network model where R is the resistance to heat exchange between outdoors and indoors, and C is the thermal capacitance of the zone. Note that this is a simpler model of building hygro-thermal dynamics than that used in the plant simulation. The constraint \eqref{eq:W_z_eq} is due to the zone humidity dynamics which is a discretized form of \eqref{eq:humidity_dynamic} presented in Section~\ref{section:hygrothermal}. 

Constraints \eqref{eq:cc_T_ca} and \eqref{eq:cc_W_ca} are for the cooling and dehumidifying coil model which is presented in the next subsection (Section~\ref{section:control_oriented_cooling_coil}).

Constraints \eqref{eq:T_z_ineq} and \eqref{eq:W_z_ineq} are thermal comfort constraints: they specify the range in which the zone temperature and humidity ratio can vary without compromising occupants' comfort. The upper and lower limits for these vary based on the scheduled hours of occupancy. Usually the limits during unoccupied mode (unocc) are relaxed when compared to the occupied mode (occ), i.e. $[T_z^{low,occ},T_z^{high,occ}]\subseteq[T_z^{low,unocc},T_z^{high,unocc}]$, $[W_z^{low,occ},W_z^{high,occ}]\subseteq [W_z^{low,unocc},W_z^{high,unocc}]$, as shown in Figure~\ref{fig:thermal_comfort_envelope}.

Constraints \eqref{eq:m_sa_ineq1} and \eqref{eq:m_sa_ineq2} are to take into account the capabilities of the fan. The minimum allowed value for the supply air flow rate is computed based on the ventilation requirements specified in ASHRAE 62.1 \cite{ASHRAE62-2016} as well as to maintain positive building pressurization. ASHRAE 62.1
demands ventilation based on two factors: number of people and floor area. Positive pressurization is required as dehumidification results in a drop in indoor vapor pressure. This negative pressure gradient may cause the infiltration of moisture from outside, especially if the building envelope is not airtight~\cite{ASHRAE_handbook_fund:17}. The minimum allowed supply air flow rate is:
\begin{align} \label{eq:62_1}
m_{sa}^{low} = max\,\,\,\big(\,\,(m_{oa}^pn_p + m_{oa}^AA)/{r_{oa}},\,\,\,\, m_{oa}^{bp}/r_{oa}\,\,\big)
\end{align}
where $m_{oa}^p$ is the outdoor air rate required per person, $n_p$ is the number of people, $m_{oa}^A$ is the outdoor air required per zone area, $A$ is the zone area, $m_{oa}^{bp}$ is the outdoor air rate required to maintain positive building pressurization, and $r_{oa}$ is the outdoor air ratio.

Constraints \eqref{eq:r_oa_ineq1}-\eqref{eq:T_sa_ineq2} are to take into account the capabilities of the damper actuators, cooling and reheat coils. In constraints \eqref{eq:T_ca_ineq1} and \eqref{eq:W_ca_ineq} the inequalities $T_{ca}(k+1)\leq T_{ma}(k+1)$ and $W_{ca}(k)\leq W_{ma}(k)$ ensure that the cooling coil can only cool and dehumidify the mixed air stream; it cannot add heat or moisture. Similarly, in constraint \eqref{eq:T_sa_ineq2} the inequality $T_{sa}(k+1)\geq T_{ca}(k+1)$ ensures that the reheat coil can only add heat; it cannot cool.

\subsubsection{Cooling and dehumidifying coil model used in $\mpcwh$}\label{section:control_oriented_cooling_coil}
Even though the binned model of cooling and dehumidifying coil presented in Section~\ref{section:cooling_coil} is quite accurate, it cannot be used in the optimizer as doing so makes the optimization a mixed integer nonlinear programming (MINLP) problem which is quite challenging to solve. Therefore, we develop a control-oriented cooling and dehumidifying coil model which makes the optimization problem a nonlinear program (NLP). It is a static model with the outputs being a polynomial function of the inputs. Note that when the chilled water flow rate is zero, no cooling or dehumidifying of the air can occur so that the conditioned air temperature and humidity ratio must be equal to the mixed air temperature and humidity ratio: $T_{ca}=T_{ma}$ and $W_{ca}=W_{ma}$, when $m_w=0$. To make the model have this behavior, the following functional form  is chosen:
\begin{align} \label{eq:control_cc}
T_{ca} &= T_{ma} + m_w \; f(T_{ma},W_{ma},m_{sa},m_w) \\
W_{ca} &= W_{ma} + m_w \; g(T_{ma},W_{ma},m_{sa},m_w)
\end{align}
For the functions $f$ and $g$, we use a quadratic form as higher degree polynomials did not show substantial gain in accuracy. The final form of the model is:
\begin{align} \label{eq:control_cc_T_ca}
T_{ca} &= f_{co}(T_{ma},W_{ma},m_{sa},m_w) \\
&= T_{ma} + m_w\big[\alpha_1 T_{ma} + \alpha_2 W_{ma} + \alpha_3 m_{sa} + \alpha_4 m_w + \alpha_5 + \nonumber\\ 
& \quad \alpha_6 T_{ma}^2 + \alpha_7 W_{ma}^2 + \alpha_8 m_{sa}^2 + \alpha_9 m_w^2 + \nonumber\\
& \quad \alpha_{10} T_{ma}W_{ma} + \alpha_{11} W_{ma}m_{sa} + \alpha_{12} m_{sa}m_w + \alpha_{13} m_wT_{ma} + \nonumber \\
& \quad \alpha_{14} T_{ma}m_{sa} + \alpha_{15} W_{ma}m_w \big] \nonumber \\
W_{ca} &= g_{co}(T_{ma},W_{ma},m_{sa},m_w) \\
&= W_{ma} + m_w\big[\beta_1 T_{ma} + \beta_2 W_{ma} + \beta_3 m_{sa} + \beta_4 m_w + \beta_5 + \nonumber\\ 
& \quad \beta_6 T_{ma}^2 + \beta_7 W_{ma}^2 + \beta_8 m_{sa}^2 + \beta_9 m_w^2 + \nonumber\\
& \quad \beta_{10} T_{ma}W_{ma} + \beta_{11} W_{ma}m_{sa} + \beta_{12} m_{sa}m_w + \beta_{13} m_wT_{ma} + \nonumber \\
& \quad \beta_{14} T_{ma}m_{sa} + \beta_{15} W_{ma}m_w \big] \nonumber,
\end{align}
where the $\alpha_i$'s and $\beta_j$'s are the model parameters to be determined. For the numerical results shown next, data obtained from EnergyPlus simulations---as explained in Section~\ref{section:cooling_coil}---are used to fit these parameters. In practice, measurements can be used to fit them.  For the validation data set, the maximum prediction errors observed are 1.61~\degree$C$ ($3~\degree F$) and $1.1\times10^{-3}$~$kg_w/kg_{da}$ for $T_{ca}$ and $W_{ca}$, respectively. This is twice the maximum error observed when using the binned cooling and dehumidifying coil model presented in Section~\ref{section:cooling_coil}. 

%\begin{figure}[htpb]
%	\subfigure[][Inputs for the binned and control-oriented cooling coil models.\label{fig:cooling_coil_plant_model_validation_inputs}]{\includegraphics[width=0.48\textwidth]{cooling_coil_plant_model_validation_inputs.pdf}} \hfill
%	\subfigure[junk][Cooling coil model predictions ($T_{ca}$ and $W_{ca}$ are the conditionaed air temperature and humidity ratio outputs from EnergyPlus simulations, $\hat T_{ca}^{bin}$ and $\hat W_{ca}^{bin}$ are the outputs computed using the binned model, and $\hat T_{ca}^{co}$ and $\hat W_{ca}^{co}$ are the outputs computed using our control-oriented model).\label{fig:mpcwhum_zone_conditions}]{\includegraphics[width=0.48\textwidth]{cooling_coil_plant_model_validation_outputs.pdf}}
%	\caption{Comparison of predictions from the binned and control-oriented cooling coil models.}
%	\label{fig:cooling_coil_binned_co}
%\end{figure}

%\begin{figure}[htpb]
%	\centering
%	\includegraphics[width=0.98\linewidth]{cooling_coil_plant_model_validation_outputs.pdf}
%	\caption{Comparison of predictions from the binned and control-oriented cooling coil models. $T_{ca}$ and $W_{ca}$ are validation data from EnergyPlus simulations. The superscript $bin$ refers to the binned model and $co$ refers to the control oriented model. \rd{PB: not sure if this plot is adding any value} }
%	\label{fig:cooling_coil_binned_co}
%\end{figure}

\subsection{Model predictive control incorporating only sensible heat ($\mpcwoh$)}\label{section:MPC_wo_hum}
This controller is similar to the one described in Section \ref{section:MPC}, with the main difference being that the moisture and latent heat of the air are not considered. The optimization problem formulation is similar to the one presented in~\cite{ma2012distributed}. 

For this controller, the vectors $x(k)$ and $v(k)$ are defined as follows: $x(k):=T_z(k)$ and $v(k):=u(k)$, where $u(k)$ is the control command vector defined in \eqref{eq:control_command}. Note how these definitions for $x(k)$ and $v(k)$ are distinct from those for $\mpcwh$.  The optimization problem at time index $j$ is:
\begin{align} \label{eq:cost_wo_hum}
\min_{V,X} \sum\limits^{j+N-1}_{k=j} \bigg[P_{fan}(k) + P_{cc}(k) + P_{reheat}(k)\bigg] \Delta t,
\end{align}
subject to the constraints: \eqref{eq:T_z_eq},\eqref{eq:T_z_ineq}, \eqref{eq:m_sa_ineq1}-\eqref{eq:T_sa_ineq2}, 
where $P_{fan}$ and $P_{reheat}$ are given by \eqref{P_fan} and \eqref{P_reheat}, and 
\begin{align} \label{eq:P_cc_wo_hum}
P_{cc}(t) = \frac{m_{sa}(t)C_{pa}\big[T_{ma}(t)-T_{ca}(t)\big]}{\eta_{cc}COP_c},
\end{align}
where $T_{ma}(t)$ and $T_{ca}(t)$ are the dry bulb temperatures of the mixed and conditioned air.

Notice the difference with $\mpcwh$: since this controller does not consider humidity and latent heat, the constraints placed on the humidity ratio at various locations in the air loop as well as the zone---\eqref{eq:W_z_eq}, \eqref{eq:W_z_ineq}, and \eqref{eq:W_ca_ineq}---are no longer used. The constraints placed on the system due to the cooling and dehumidifying coil model---\eqref{eq:cc_T_ca} and \eqref{eq:cc_W_ca}---are also not present. The cooling power term in the objective function is based only on the sensible heat; latent heat is ignored.

%, $C_{pa}$ is the specific heat of air, $m_{sa}$ is the mass flow rate of supply air, $\eta_{cc}$ is the cooling coil efficiency, and $COP_c$ is the chiller coefficient of performance.

%We can see that \eqref{eq:P_cc_wo_hum} only accounts for the sensible thermal power component of the cooling coil and does not consider the latent component. Therefore, the cooling power consumption predicted by \eqref{eq:P_cc_wo_hum} will always be lower than or equal to the true power consumption of the cooling coil.

%The constraint \eqref{eq:T_z_eq_wo_hum}, similar to \eqref{eq:T_z_eq}, is due to the thermal dynamics of the zone. Constraint \eqref{eq:T_z_ineq_wo_hum} specifies the range in which the zone temperature can vary so that the thermal comfort of the occupants is satisfied. Constraints \eqref{eq:m_sa_ineq1_wo_hum} and \eqref{eq:m_sa_ineq2_wo_hum} take in to account the capability of the fan to vary air flow rates. Constraints \eqref{eq:r_oa_ineq1_wo_hum}-\eqref{eq:T_sa_ineq2_wo_hum} represent the capabilities of the damper actuators, cooling and reheat coils. Constraint \eqref{eq:T_ca_ineq1_wo_hum} ensures the cooling coil can only cool the mixed air stream. Similarly, constraint \eqref{eq:T_sa_ineq2_wo_hum} ensures that the reheat coil can only add heat to the conditioned air.

\subsection{Baseline control ($\base$)}
\begin{figure}[ht!]
	\centering
	\includegraphics[width=0.98\linewidth]{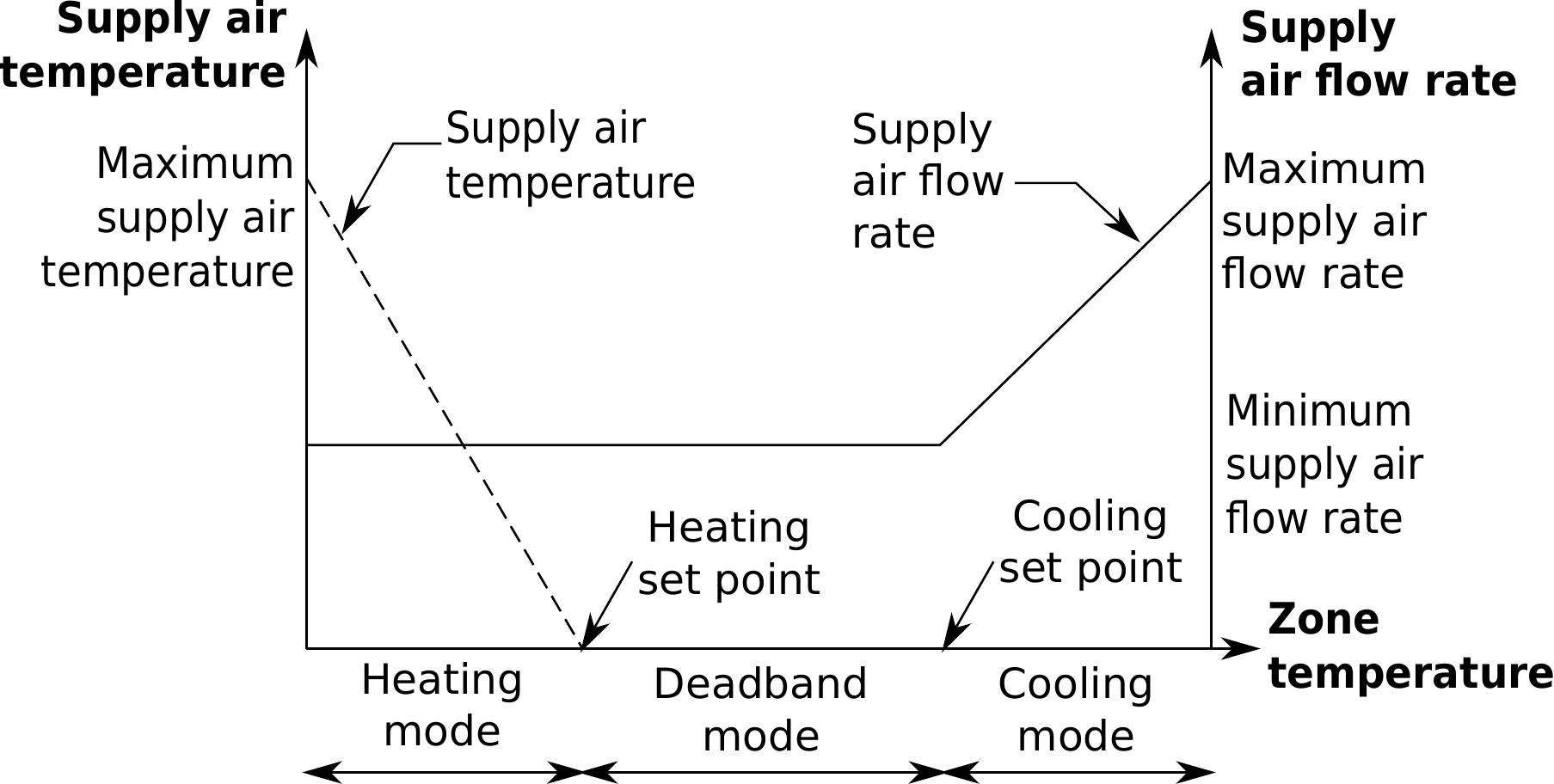}
	\caption{Schematic of Single Maximum control algorithm.}
	\label{fig:single_maximum}
\end{figure}

The baseline controller---against which the performance of the proposed $\mpcwh$, and $\mpcwoh$ is compared---is chosen to be the single maximum controller that is widely used in practice~\cite{ASHRAE_handbook_applications:11}.  In single maximum control, whose schematic representation is shown in Figure~\ref{fig:single_maximum}, the HVAC system operates in three modes based on the zone temperature: cooling, heating, and deadband. When the zone temperature is above the cooling set point for more than 5 minutes the system is in cooling mode and the supply air flow rate ($m_{sa}$) is varied between the minimum and maximum allowed values to maintain the zone temperature. Similarly, when the zone temperature is below the heating set point for more than 5 minutes the supply air temperature ($T_{sa}$) is varied to maintain the zone temperature. In this mode the supply air flow rate is kept at the minimum allowed value. Finally when the temperature is between the cooling and heating set points the system is in deadband mode with the supply air flow rate kept at the minimum and the supply air temperature is equal to the conditioned air temperature ($T_{sa}$=$T_{ca}$). The minimum allowed value for the supply air flow rate should satisfy the following conditions: one, the ventilation requirements specified by ASHRAE 62.1 \cite{ASHRAE62-2016} and positive building pressurization, as described in Section~\ref{section:MPC}. Two, it should be high enough to meet the design heating load at a supply air temperature that is low enough to prevent stratification (e.g., $30~\degree C$). The outdoor air ratio and the conditioned air temperature are kept constant at all times.

\section{Simulation setup}\label{sec:sim_setup}
The plant is simulated in SIMULINK. The optimization problem is solved using CasADi \cite{Andersson2018} and IPOPT \cite{wacbie:2006}, a nonlinear programming (NLP) solver, on a Desktop Linux computer with 16GB RAM and a 3.60 GHz $\times$ 8 CPU. On an average it takes 2~seconds for $\mpcwh$ and 0.6~seconds for $\mpcwoh$ to solve their respective optimization problems. The higher computation time for $\mpcwh$ is attributed to the larger number of decision variables. Both the NLPs are non-convex, and the NLP solver indicates that it is able to find a local minimum successfully 100$\%$ of the time. In cases where they may not be feasible, the controllers are programmed to use the control command computed at the previous time step. 

Three types of outdoor weather conditions are tested: hot-humid (Aug/06/2016), mild (Mar/25/2016), and cold (Dec/20/2016), all for Gainesville, FL, USA. The weather data is obtained from \url{www.wunderground.com} and National Solar Radiation Database (\url{nsrdb.nrel.gov}). The simulations are run for 24 hours starting at 8:00~AM.

\subsection{Plant parameters and thermal comfort envelope}\label{section:thermal_comfort}
\begin{figure}[htpb]
	\centering
	\includegraphics[width=0.98\linewidth]{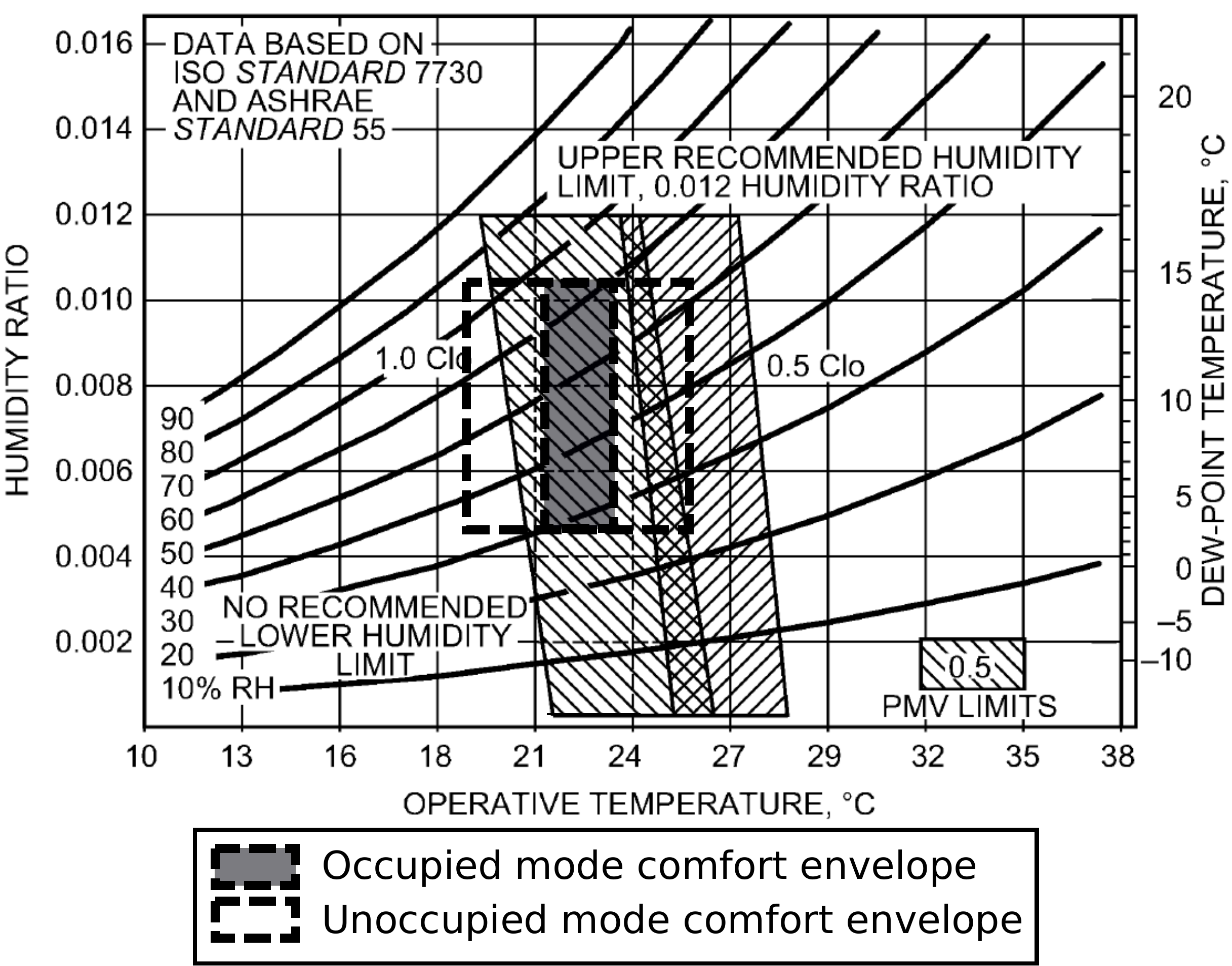}
	\caption{Thermal comfort envelope from \cite{ASHRAE_handbook_fund:17} shown as the hatched areas. Comfort envelope chosen in this paper shown as the shaded area during scheduled hours of occupancy and the unshaded area enclosed by dashed line during unoccupied hours.}
	\label{fig:thermal_comfort_envelope}
\end{figure}

The plant parameters are chosen based on a large classroom/auditorium ($\sim 6$ $m$ high, floor area of $\sim 465$ $m^2$) in Pugh Hall located in the University of Florida campus. The RC network parameters are chosen to be $R_z=0.6\times10^{-3}~\degree C/W$, $R_w=0.55\times10^{-3}~\degree C/W$, $C_z=3.132\times10^7~J/\degree C$, $C_w=7.092\times10^7~J/\degree C$, and $A_e=8.12~m^2$ from \cite{cofbar:2018}, which were obtained by fitting the model to measured data from the building. The scheduled occupancy is between 8:00 AM to 5:00 PM during which the following constraints are used: $T_z^{low,occ}=21.1~\degree C$ $(70~\degree F)$, $T_z^{high,occ}=23.3~\degree C$ $(74\degree F)$, $W_z^{low,occ}=0.0046~kg_w/kg_{da}$, and $W_z^{high,occ}=0.0104~kg_w/kg_{da}$. The unoccupied hours are between 5:00 PM to 8:00 AM during which the constraints are: $T_z^{low,unocc}=18.9~\degree C$~$(66~\degree F)$, $T_z^{high,unocc}=25.6~\degree C$~$(78~\degree F)$, $W_z^{low,unocc}=0.0046~kg_w/kg_{da}$, and $W_z^{high,unocc}=0.0104~kg_w/kg_{da}$. The chosen thermal comfort envelope is shown in Figure~\ref{fig:thermal_comfort_envelope}. 

The values for $m_{oa}^p=0.0043~kg/s/person$ $(7.5~cfm/person)$ and $m_{oa}^A=3.67\times10^{-4}~kg/s/m^2$ $(0.06~cfm/ft^2)$ are chosen based on ASHRAE 62.1~\cite{ASHRAE62-2016} for a lecture classroom. For positive pressurization, $m_{oa}^{bp}=0.1894~kg/s$ is chosen so that there are 0.2 air changes per hour.  $q_{ocp}$ and $\omega_{ocp}$ are computed based on the number of people in the zone, assuming that each person produces $100~W$ of heat and $1.39\times10^{-5}~kg/s$ ($50~g/hr$) of water vapor \cite{ASHRAE_handbook_fund:17}, with $n_p$ being 175, which is the design occupancy for the building. $q_{other}$ is chosen to be $6000~W$ based on lighting/equipment power density of $12.92~W/m^2$ ($1.2~W/ft^2$) and $\omega_{other}$ is $0~kg/s$.

\subsection{Controller parameters}
The controller parameters for $\mpcwh$ and $\mpcwoh$ are listed in Table~\ref{table:sim_para}. For the  1R-1C model used in the $\mpcwh$ and $\mpcwoh$, we use $R=1.15\times10^{-3}~\degree C/W$ and $C=6.0167\times10^7~J/\degree C$. These values are obtained by creating a 1R-1C model equivalent to the 2R-2C model \eqref{eq:thermal_dynamic}, and equating the DC gains and rise times for the transfer functions, with $T_{oa}$ and the heat gains as inputs and the zone temperature as output. As mentioned earlier, $\Delta t = 5$ minutes and $N=288$ (i.e., 24 hours). The number of decision variables for $\mpcwh$ is $2304$ and $\mpcwoh$ is $1440$.

\begin{table}[h]
	\caption{MPC parameters.}
	\label{table:sim_para}
	\begin{center}
		\begin{tabular}{|c|c|c|c|c|}
			\hline
			$~$ & $~$ & $~$ & $~$ & $~$ \\
			$m_{sa}^{high}$ & $T_{ca}^{low}$ & $T_{sa}^{high}$ & $r_{oa}^{low}$ & $r_{oa}^{high}$ \\
			$(kg/s)$ & $(\degree C)$ & $(\degree C)$ & $(\%)$ & $(\%)$ \\
			$4.6$ & $12.8$ & $30$ & $0$ & $100$ \\ 
			\hline
			$~$ & $~$ & $~$ & $~$ & $~$ \\
			$m_{sa}^{rate}$ & $T_{ca}^{rate}$ & $T_{sa}^{rate}$ & $r_{oa}^{rate}$ & $\Delta t$ \\
			$(kg/s/min)$ & $(\degree C/min)$ & $(\degree C/min)$ & $(\%/min)$ & $(min)$ \\
			$0.37$ & $0.56$ & $0.56$ & $6$ & $5$\\
			\hline
		\end{tabular}
	\end{center}
\end{table}

For the baseline controller, outdoor air ratio is kept at $30\%$ and conditioned air temperature is kept at $12.8~\degree C$ ($55~\degree F$).

\subsection{Performance metrics}\label{sec:performance_metrics}
To evaluate the various controllers in this study, we look at the energy consumed by each of them as well as the violations caused with respect to thermal comfort limits specified in Section~\ref{section:thermal_comfort}.

The total energy consumed by the controllers for 24 hours is computed as follows:
\begin{align}
E_{total} &= \int_{24hrs} P_{fan}(t) + P_{cc}(t) + P_{reheat}(t) \,\,dt,
\end{align}
where $P_{fan}$, $P_{cc}$, and $P_{reheat}$ are computed using \eqref{P_fan}, \eqref{P_cc}, and \eqref{P_reheat} respectively.

We define the daily temperature violation as:
\begin{align} \label{V_T}
V_T  &= \int_{24\,hrs} \Delta T_z(t)dt,
\end{align}
where the term $\Delta T_z(t)$ is defined as \cite{SG_HI_PB_AE:2013}:
\begin{align} \label{Delta_T_z}
\Delta T_z(t) &= \begin{cases}
T_z(t) - T_z^{high}, &\text{ if } T_z(t)>T_z^{high} \\
T_z^{low} - T_z(t), &\text{ if } T_z(t)<T_z^{low} \\
0, & \text{ otherwise}.
\end{cases}
\end{align}
The unit of $V_T$ is $\degree C$-hours. Similarly, we define the daily humidity violation as:
\begin{align} \label{V_W}
V_W  &= \int_{24\,hrs} \Delta W_z(t)dt,
\end{align} 
where the term $\Delta W_z(t)$ is defined as \cite{SG_HI_PB_AE:2013}:
\begin{align} \label{Delta_W_z}
\Delta W_z(t) &= \begin{cases}
W_z(t) - W_z^{high}, &\text{ if } W_z(t)>W_z^{high} \\
W_z^{low} - W_z(t), &\text{ if } W_z(t)<W_z^{low} \\
0, &\text{ otherwise}.
\end{cases}
\end{align} 
The unit of $V_W$ is $kg_w/kg_{da}$-hours.

The larger $V_T$ and $V_W$ are, greater the adverse impact on occupants' comfort and health. 

\section{Results and discussions} \label{sec:results_discussions}
\subsection{Results for the different outdoor weather conditions}
\subsubsection{Hot-humid day}
\begin{figure*}[!ht]
	%\captionsetup{justification=centering}
	\centering
	\subfigure[Outdoor weather data and occupancy profile used in simulations (outdoor air temperature, outdoor air relative humidity, solar irradiance, and number of people).\label{fig:outdoor_weather_hot}]{\includegraphics[width=0.49\textwidth]{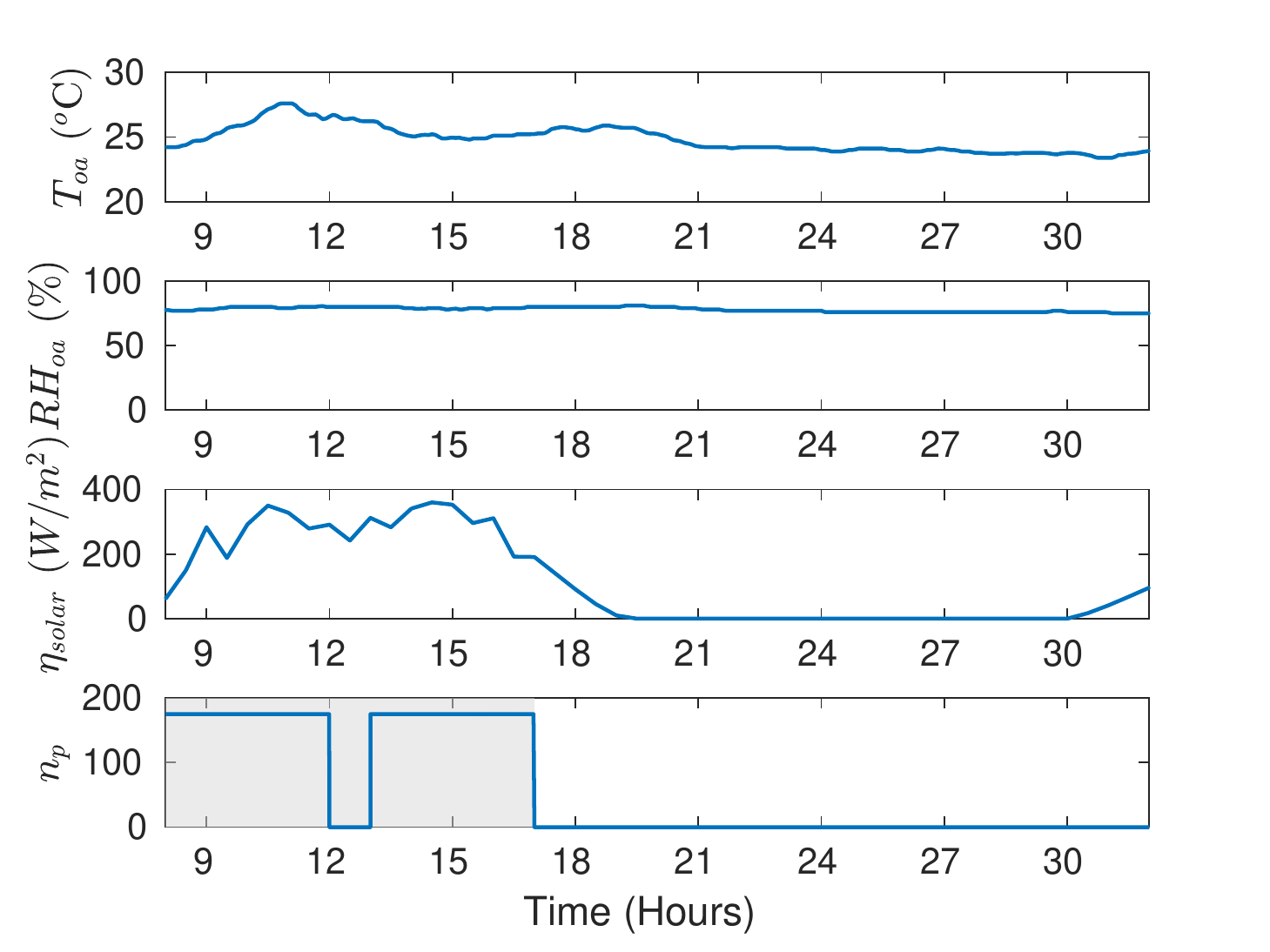}}\quad
	\subfigure[Comparing the power consumptions (fan, cooling, and reheat power).\label{fig:comparison_power_hot}]{\includegraphics[width=0.49\textwidth]{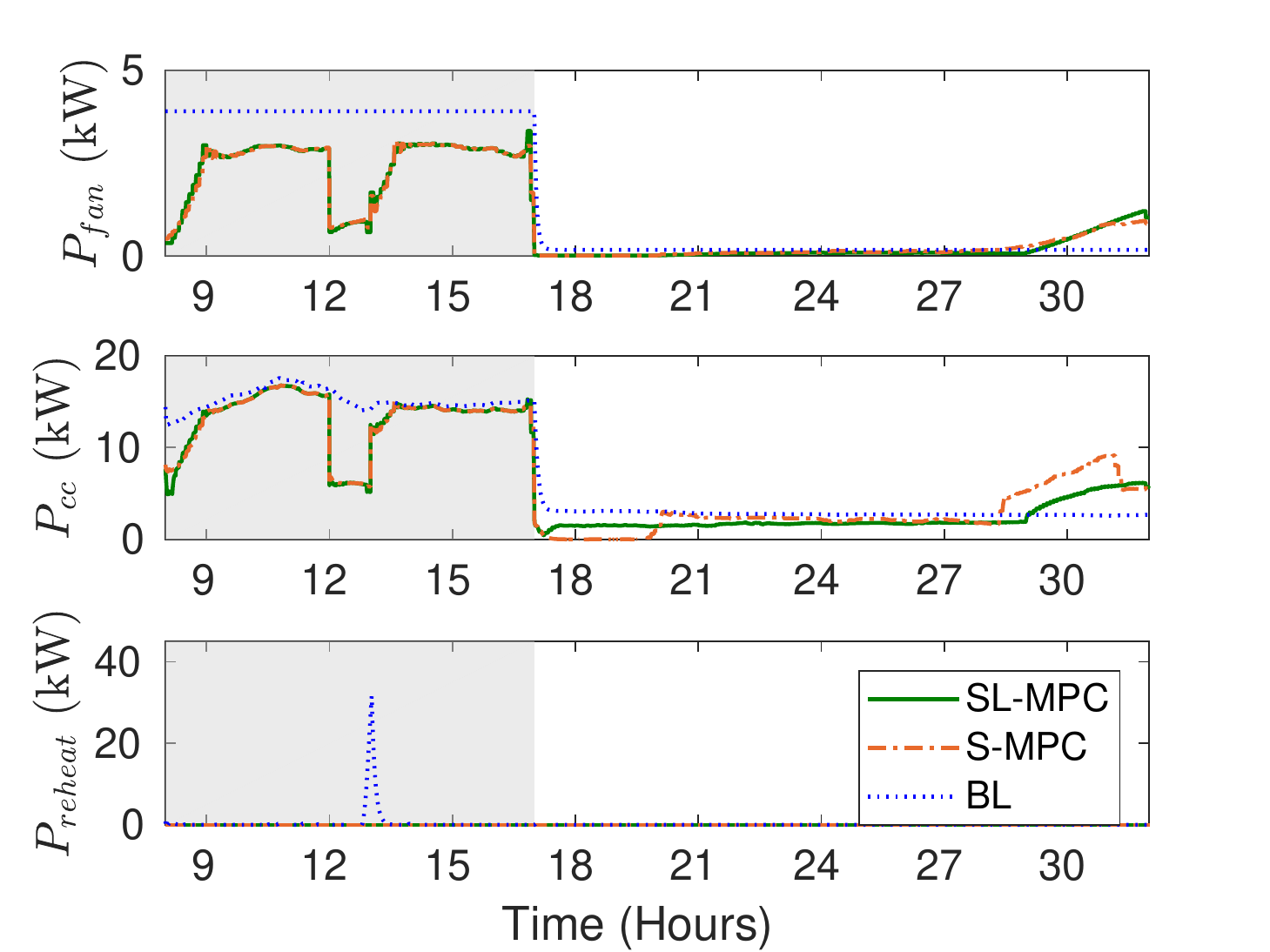}}
	%	\caption{Comparison of MPC vs Baseline simulation results.}\label{fig:mpc_simulation_results_1}
	\subfigure[Zone and air loop conditions with the black dashed lines showing the upper and lower comfort limits (zone air temperature, zone air humidity ratio, outdoor air ratio, and supply air flow rate).\label{fig:comparison_zone_conditions_1_hot}]{\includegraphics[width=0.49\textwidth]{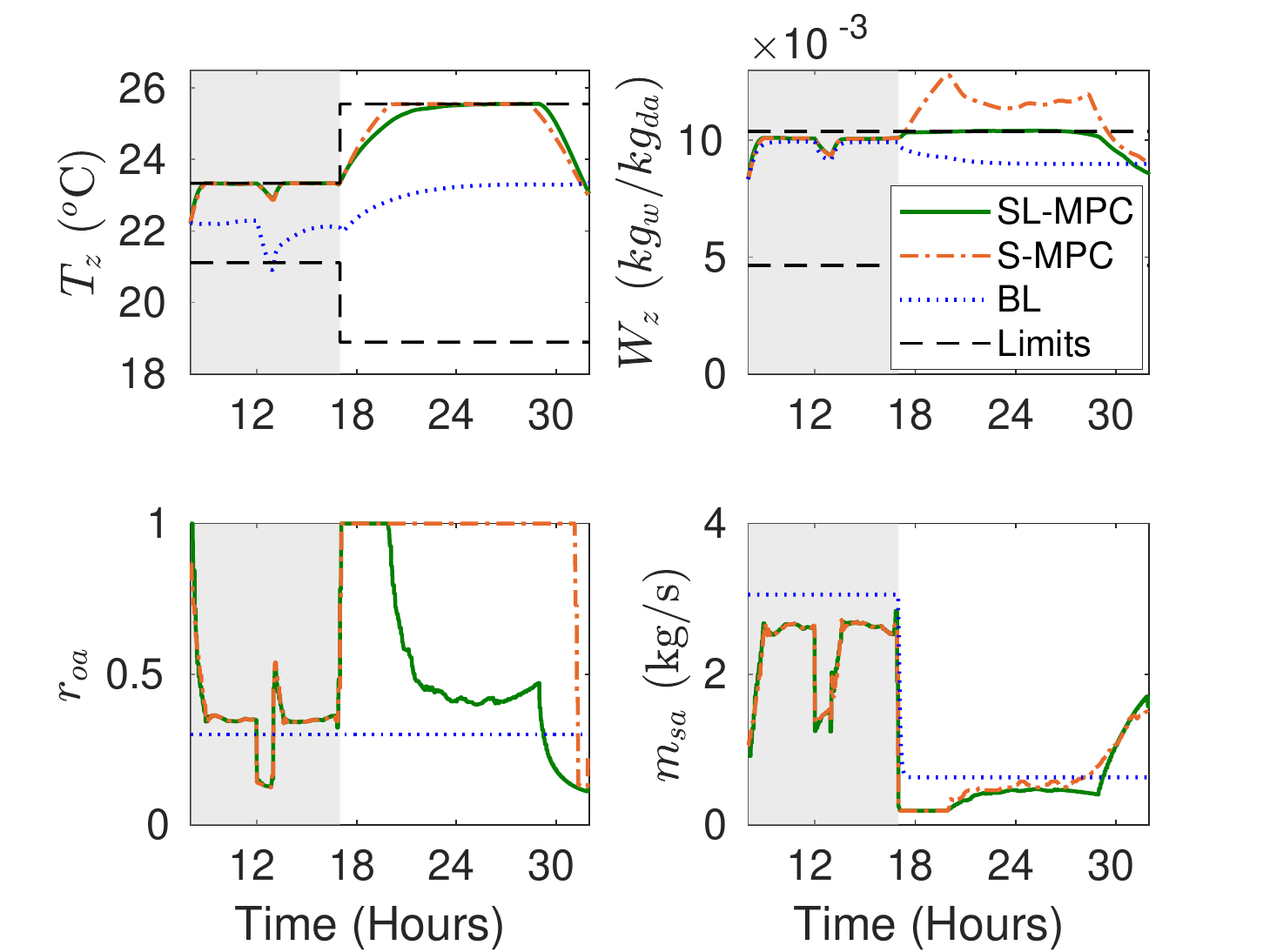}}\quad
	\subfigure[HVAC system conditions (conditioned air temperature, conditioned air humidity ratio, supply air temperature, and chilled water flow rate).\label{fig:comparison_zone_conditions_2_hot}]{\includegraphics[width=0.49\textwidth]{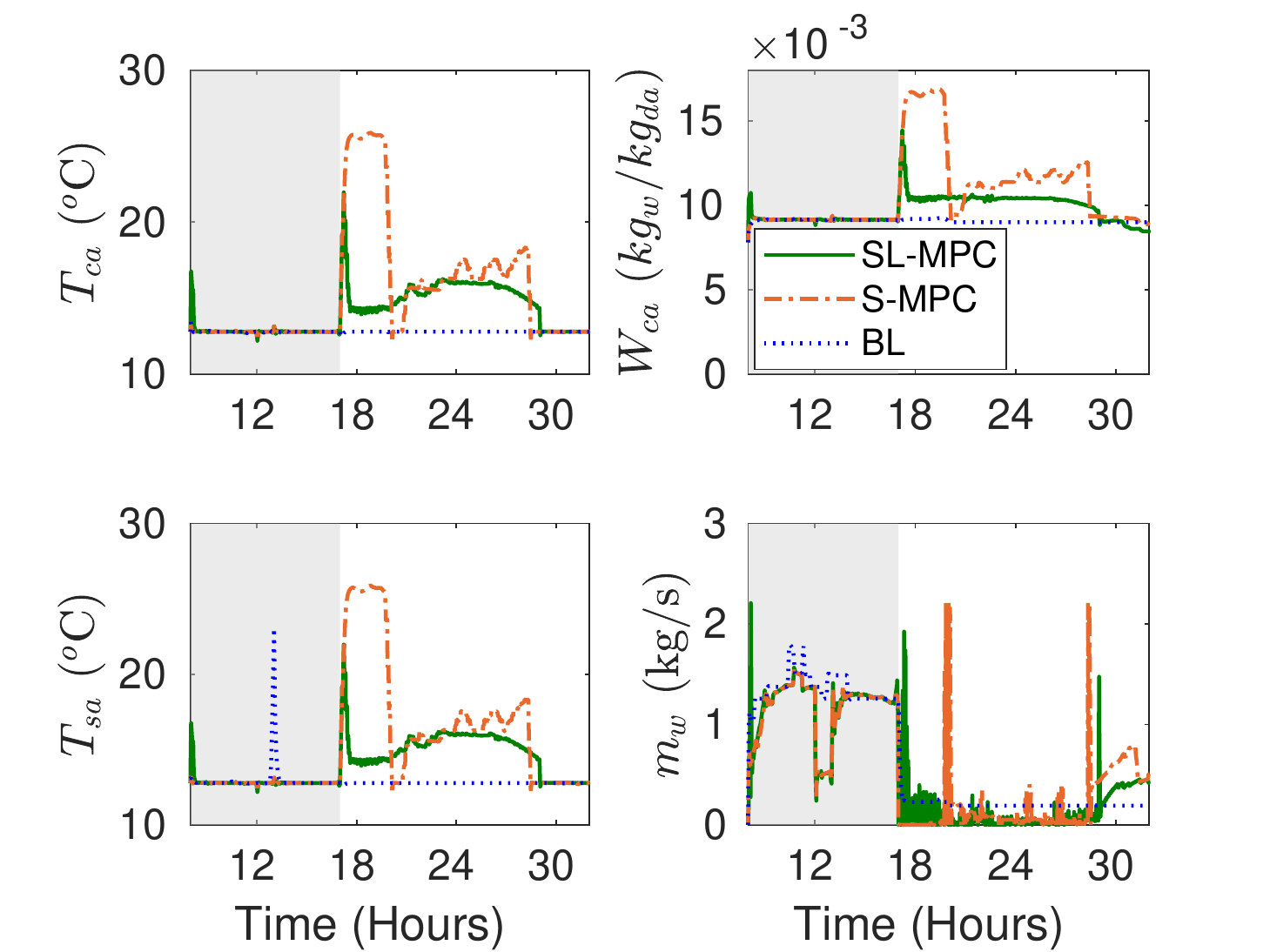}}
	\caption{Comparison of the three controllers for a hot-humid day (August/06/2016, Gainesville, Florida). The scheduled hours of occupancy are shown as the gray shaded area.}\label{fig:mpc_simulation_results_hot}
\end{figure*}

\begin{figure*}[!ht]
	%\captionsetup{justification=centering}
	%\centering
	\subfigure[Outdoor weather data and occupancy profile used in simulations (outdoor air temperature, outdoor air relative humidity, solar irradiance, and number of people).\label{fig:outdoor_weather_mild}]{\includegraphics[width=0.49\textwidth]{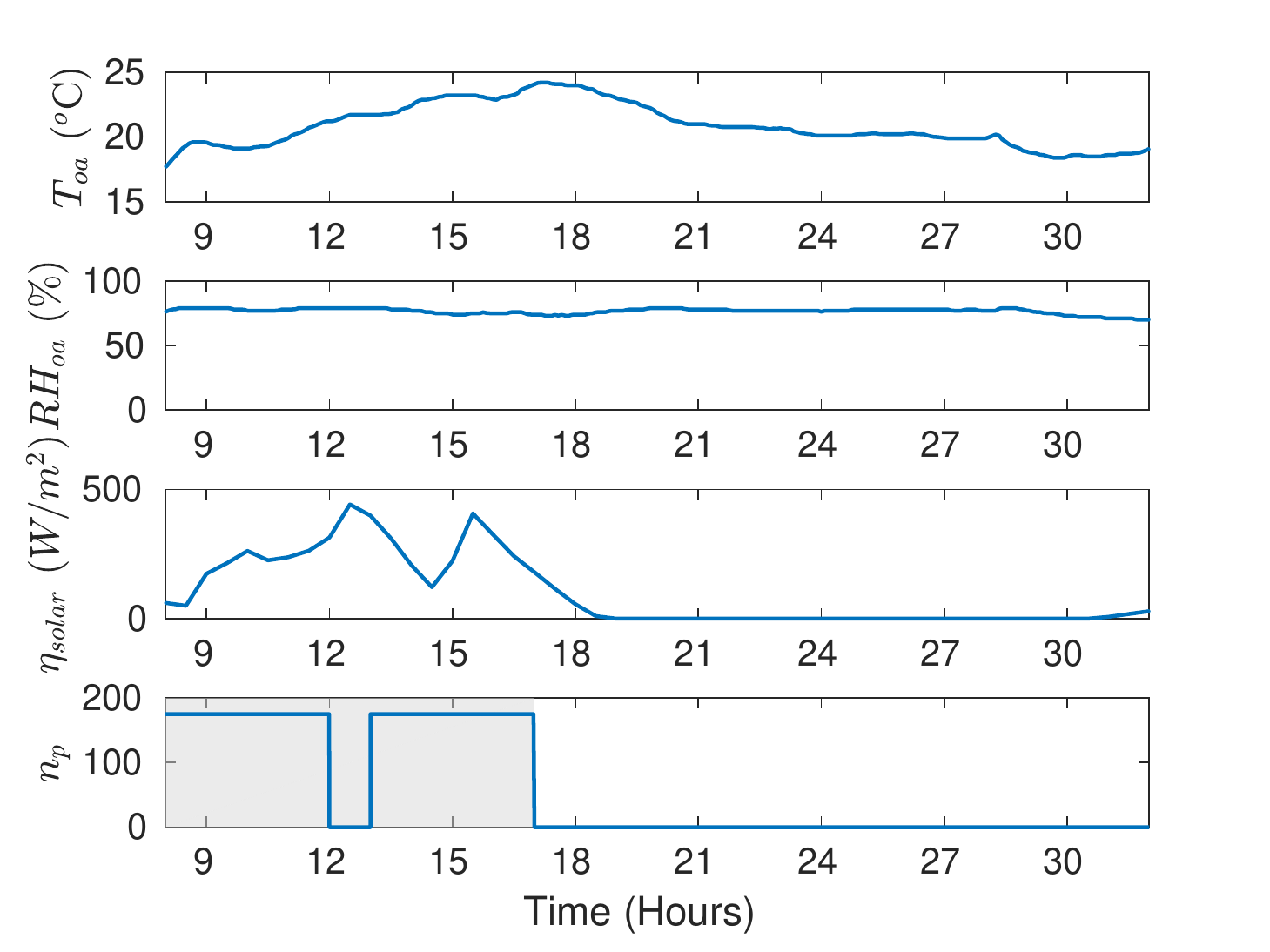}}\quad
	\subfigure[Comparing the power consumptions (fan, cooling, and reheat power).\label{fig:comparison_power_mild}]{\includegraphics[width=0.49\textwidth]{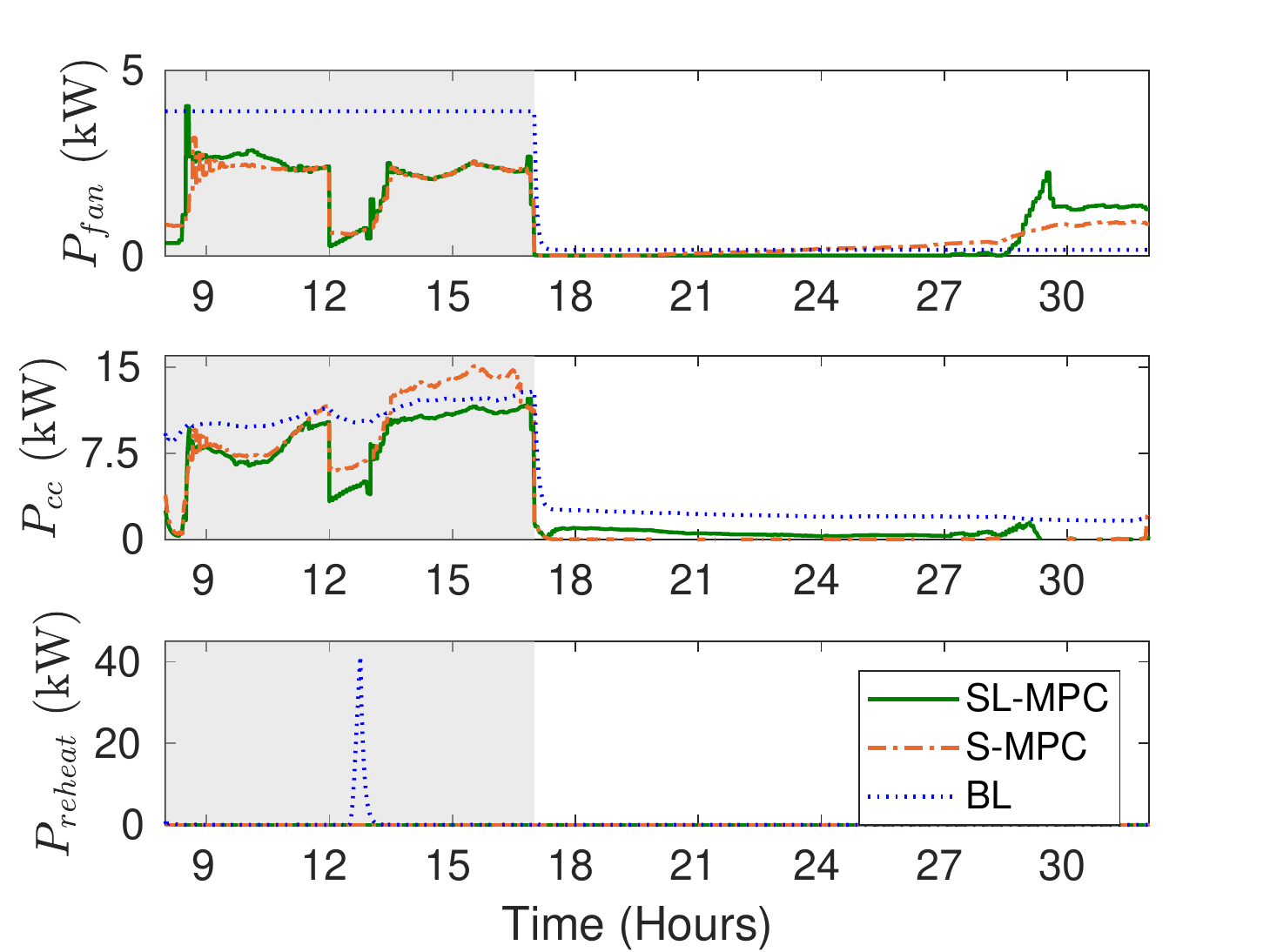}}
	%	\caption{Comparison of MPC vs Baseline simulation results.}\label{fig:mpc_simulation_results_1}
	
	\subfigure[Zone and air loop conditions with the black dashed lines showing the upper and lower comfort limits (zone air temperature, zone air humidity ratio, outdoor air ratio, and supply air flow rate).\label{fig:comparison_zone_conditions_1_mild}]{\includegraphics[width=0.49\textwidth]{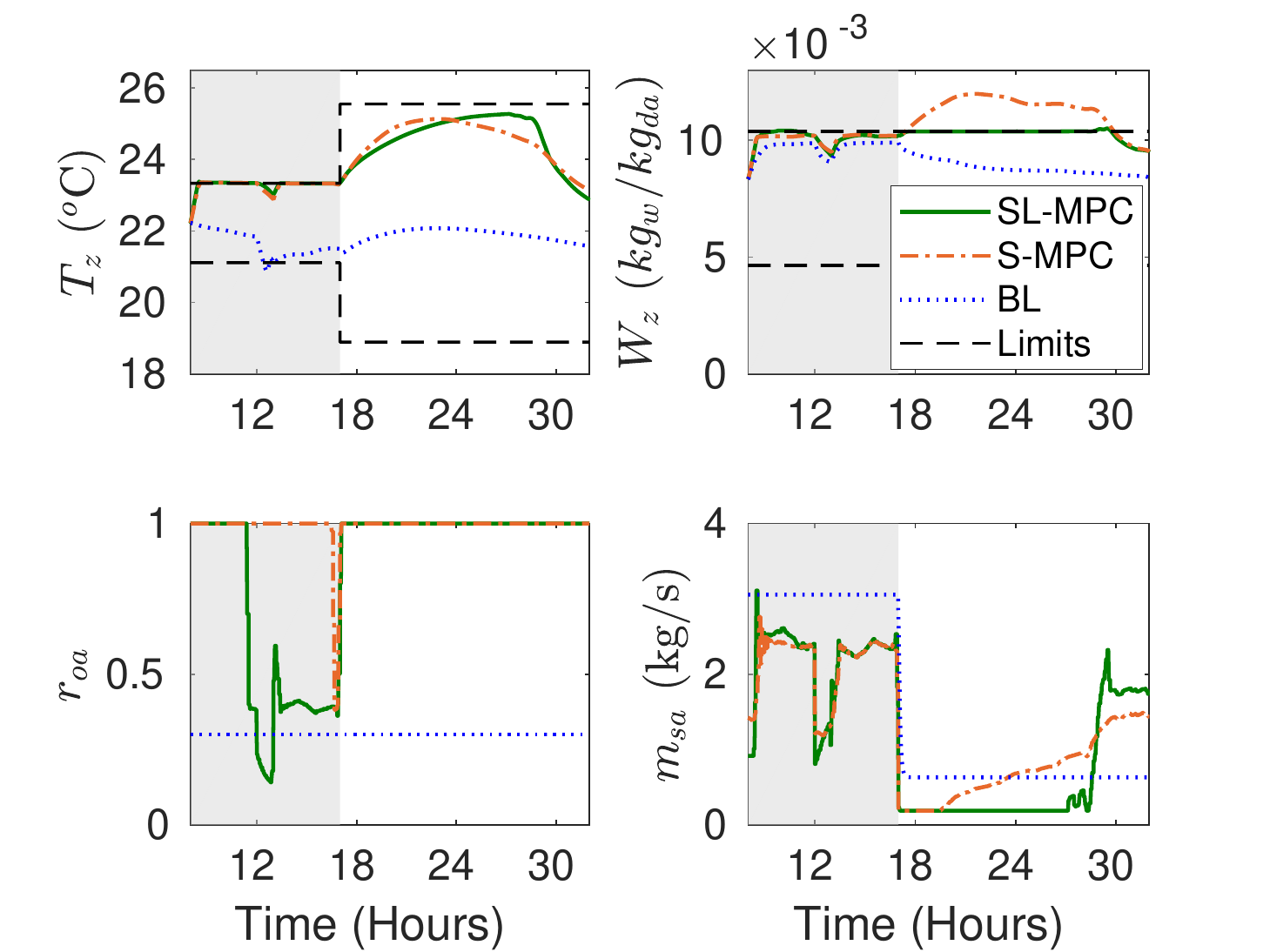}}\quad
	\subfigure[HVAC system conditions (conditioned air temperature, conditioned air humidity ratio, supply air temperature, and chilled water flow rate).\label{fig:comparison_zone_conditions_2_mild}]{\includegraphics[width=0.49\textwidth]{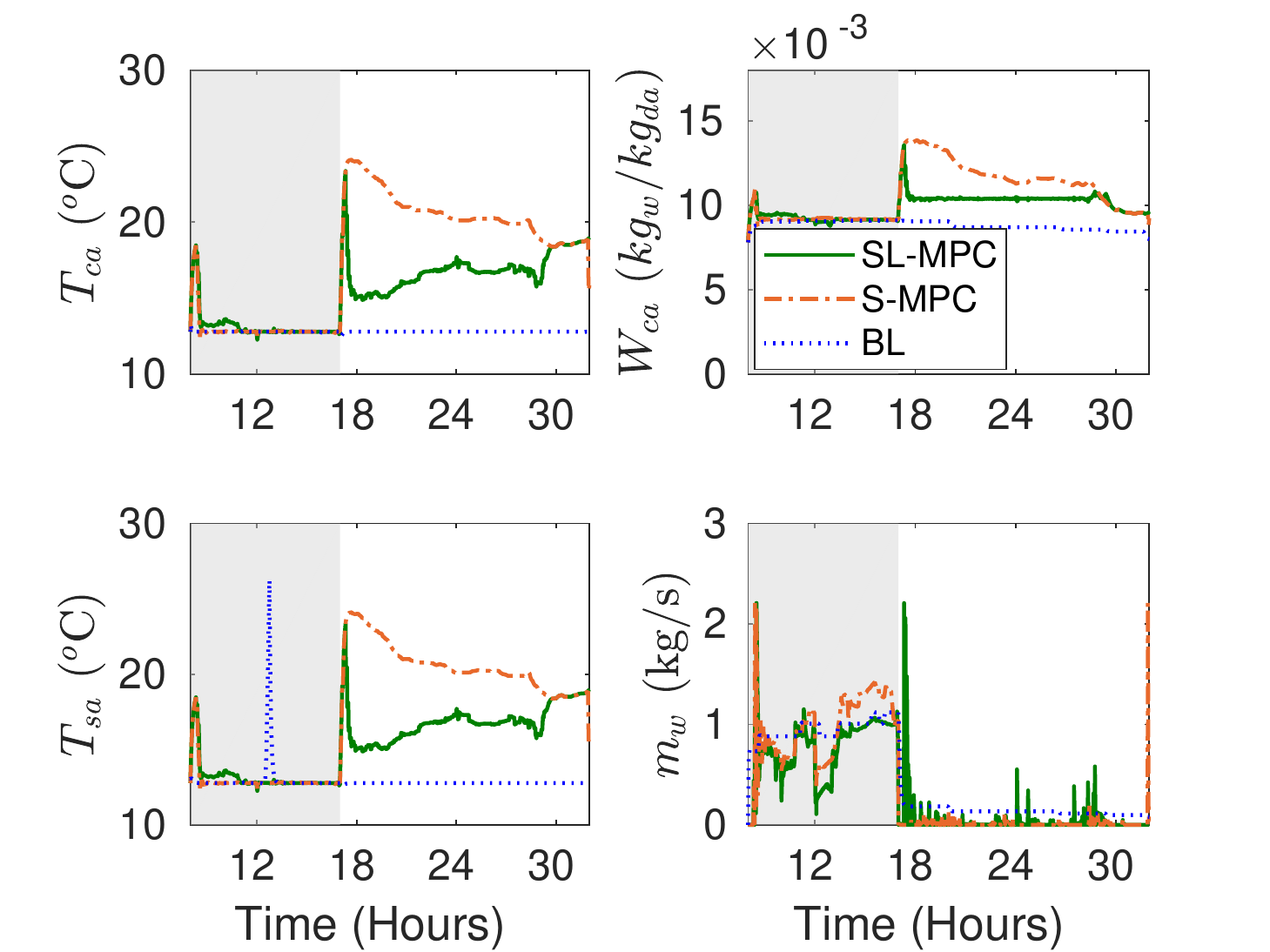}}
	\caption{Comparison of the three controllers for a mild day (March/25/2016, Gainesville, Florida). The scheduled hours of occupancy are shown as the gray shaded area.}\label{fig:mpc_simulation_results_mild}
\end{figure*}
Figure~\ref{fig:mpc_simulation_results_hot} shows the simulation results for a hot-humid day. It is found that $\mpcwh$ consumes the least amount of energy when compared to $\mpcwoh$ and $\base$, as presented in Figure~\ref{fig:performance_metrics}. There are large violations in humidity limits by $\mpcwoh$, specifically during the unoccupied hours, as shown in Figures~\ref{fig:performance_metrics} and \ref{fig:comparison_zone_conditions_1_hot}.

All three controllers are able to maintain thermal comfort limits during scheduled hours of occupancy almost all the time; see 08:00-17:00 hours in Figure~\ref{fig:comparison_zone_conditions_1_hot}. The $\base$ ensures that dry air is supplied to the zone and hence the humidity limit is not violated since it keeps the conditioned air temperature at a constant value of 12.8~$\degree C$ (55$\degree F$). In the case of $\mpcwoh$, the optimal control decisions made by it are observed to be similar to those made by $\mpcwh$. This can be attributed to the high internal heat load and hot outdoor air temperature. Specifically, $\mpcwoh$ decides to keep the conditioned air temperature low enough (at $12.8~\degree C$) to meet the heat load which has the unintended, but good, side effect of maintaining zone humidity within the comfort limits.

Both the MPC controllers consume lesser energy when compared to $\base$ during occupied hours; see 08:00-17:00 hours in Figure~\ref{fig:comparison_power_hot}. The reason for this is that the outdoor air ratio is kept constant for $\base$ and it also assumes full occupancy from 08:00 to 17:00 hours. Therefore, the air flow rate has to be kept high enough so that the ventilation requirements specified in ASHRAE 62.1~\cite{ASHRAE62-2016} are met. This high air flow rate, combined with the low conditioned air temperature, is highly suboptimal, especially when there is a reduction in occupancy: not only is the air cooled unnecessarily, but reheating must also be performed to prevent the zone from becoming too cold. This phenomenon can be seen between 12:00-13:00 hours in Figures~\ref{fig:comparison_power_hot} and \ref{fig:comparison_zone_conditions_1_hot}. The MPC controllers in contrast vary the outdoor air ratio and air flow rate as occupancy varies, leading to a lower fan and cooling energy consumption. It should be noted that this reduction in energy use by the MPC controllers requires accurate prediction of occupancy.

From Figure~\ref{fig:comparison_zone_conditions_1_hot} it can be seen that $\mpcwoh$ violates the humidity limits during unoccupied hours while $\mpcwh$ and $\base$ do not. This is because $\mpcwoh$ decides to bring in the slightly cooler outside air in an attempt to provide ``free'' cooling but fails to realize that the air is humid. If this violation of humidity limit occurs over several months, serious and costly issues such as mold growth are a real possibility~\cite{baughman1996indoor}.This does not occur with $\mpcwh$ as humidity is a part of the problem formulation---the humidity constraint is found to be active between 18:00-28:00 hours as shown in Figure~\ref{fig:comparison_zone_conditions_1_hot}.

The difference in energy consumption between $\mpcwoh$ and $\mpcwh$ occurs over unoccupied hours. This is another effect of the attempt to use ``free'' cooling by $\mpcwoh$. The use of slightly cool, but humid, outdoor air results in the cooling coil always having to reduce the temperature of mixed air and de-humidify it, resulting in high power consumption. In the case of $\mpcwh$, it decides to re-circulate return air and reduce the outdoor air ratio, thereby reducing the amount of de-humidification required. This lowers the cooling coil energy consumption and the overall energy consumption.

%An interesting observation is that when the heat load is high, the optimal value for $T_{ca}$ found by $\mpcwh$ is 285.93~K (55$\degree F$, see Figure~\ref{fig:comparison_zone_conditions_2_hot} between 9:00-17:00 hours). Incidentally, this is a widely used rule of thumb, especially in hot-humid climates.

\subsubsection{Mild day}
Figure~\ref{fig:mpc_simulation_results_mild} shows the simulation results for a mild day. It is found that $\mpcwh$ consumes the least amount of energy when compared to $\mpcwoh$ and $\base$, as seen in Figure~\ref{fig:performance_metrics}. Similar to the results from hot day, there are huge violations in humidity limit during unoccupied hours when using $\mpcwoh$. The conservative set points in $\base$ ensure that the comfort limits are not violated but at the cost of high energy use. Therefore, we discuss only the MPC controllers in further detail here.

As discussed in Section~\ref{sec:system_models}, there are four control commands the MPC controllers need to decide. They are $m_{sa}$, $r_{oa}$, $T_{ca}$, and $T_{sa}$. Since the weather condition is not too cold, there will be no reheat ($T_{sa}=T_{ca}$) and the controllers need to decide the remaining three: $m_{sa}$, $r_{oa}$, and $T_{ca}$. During the occupied hours, it is seen that $\mpcwoh$ decides to keep $T_{ca}$ low enough (at $12.8~\degree C$) similar to $\mpcwh$ due to the high internal heat load, and hence maintains the zone humidity. This behavior is similar to the one seen for a hot-humid day. But, the biggest difference in the decisions made by the two MPC controllers are for $r_{oa}$ and $m_{sa}$. $\mpcwoh$ decides to use $100\%$ of the slightly cold outside air in an attempt to lower the cooling and fan energy consumption, but fails to realize that it is humid. Whereas $\mpcwh$ uses much lesser outside air. As a result, the cooling energy consumed by $\mpcwoh$ is much higher than that consumed by $\mpcwh$, and can be seen between 12:00-17:00 hrs in Figures~\ref{fig:comparison_power_mild} and \ref{fig:comparison_zone_conditions_1_mild}.

During unoccupied hours both the MPC controllers decide to bring in 100$\%$ outside air, but $\mpcwh$ decides to keep $T_{ca}$ lower than the one decided by $\mpcwoh$ to ensure that the air is dehumidified enough before being supplied to the zone.

\subsubsection{Cold day}
Since the outdoor weather is dry, no matter what decisions are made by a controller, it is unlikely to violate humidity constraints in the building. The energy consumed by the two MPC controllers is almost the same, which is much smaller than that by $\base$ (Figure~\ref{fig:performance_metrics}). $\base$ performs simultaneous heating and cooling in a pronounced manner leading to high energy consumption: the fixed outdoor air ratio combined with the 12.8~\degree$C$ (55~\degree $F$) conditioned air requires usage of cooling energy, additionally reheating is required to keep the building warm enough because of the cold weather. %It is found that a simple change of $r_{oa}$ from a constant value of 30$\%$---which is the value used when computing $E_{total}$ in Figure~\ref{fig:performance_metrics}---to 40$\%$ cuts down the total energy consumed by $\base$ by more than half. 
The MPC controllers choose to use as much outdoor air as possible, since the cold outdoor conditions provide free cooling without having to use chilled water. %Increasing the outdoor air ratio reduces the required supply air flow rate, leading to a lower fan and cooling energy consumption.

\subsection{Comparison among controllers}
\begin{figure}[!ht]
	\centering
	\subfigure[][Comparison of total energy consumed over 24 hours by $\mpcwh$, $\mpcwoh$, and $\base$ for different outdoor weather conditions.\label{fig:total_energy_consumption}]{\includegraphics[width=0.44\textwidth]{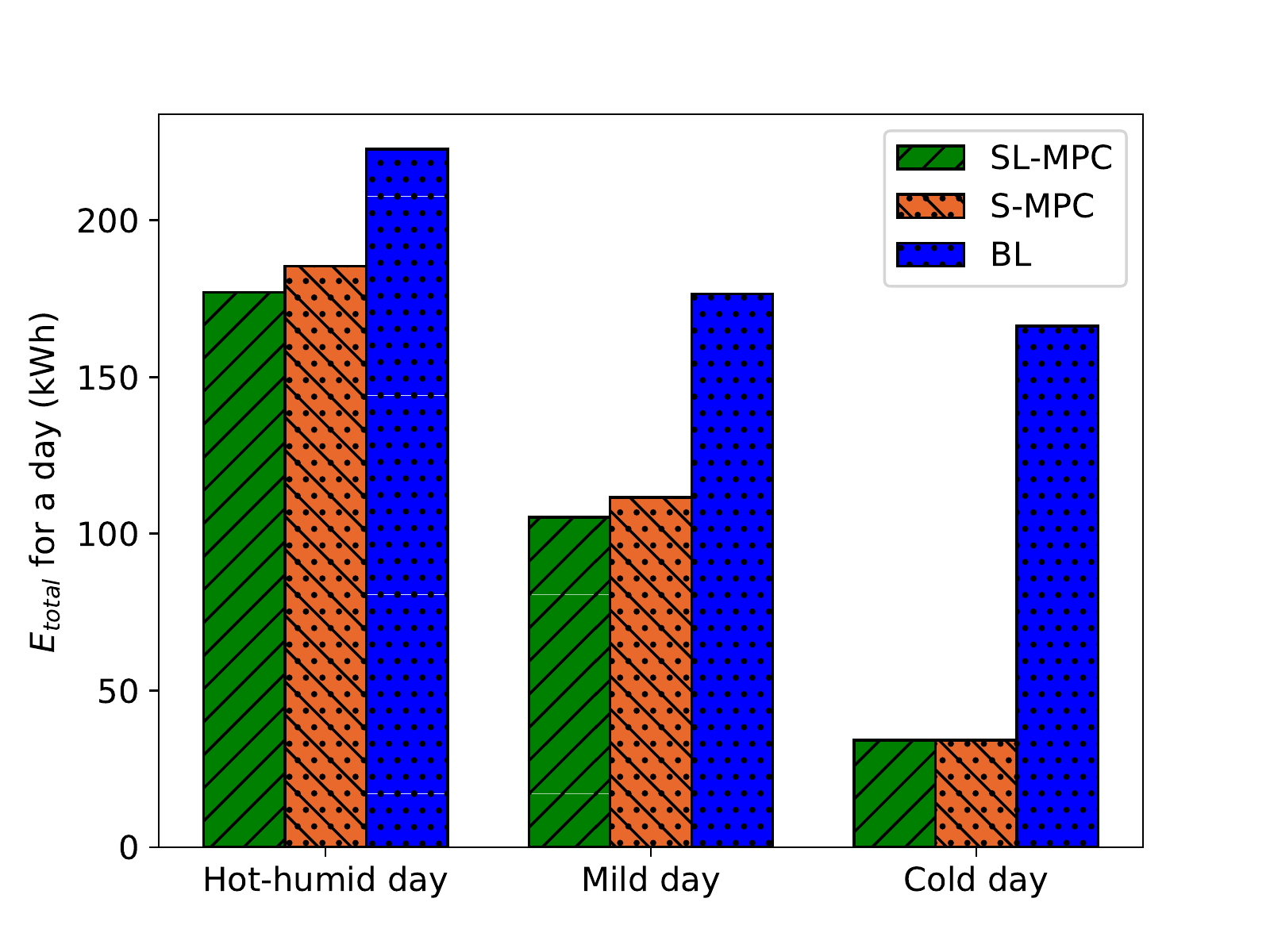}} \hfill
	\subfigure[junk][Comparison of humidity violation over 24 hours by $\mpcwh$, $\mpcwoh$, and $\base$ for different outdoor weather conditions. The black line indicates the value of $V_W$ when there is a 2.5$\%$ violation over the upper limit of humidity ratio at every instant. \label{fig:humidity_violation}]{\includegraphics[width=0.44\textwidth]{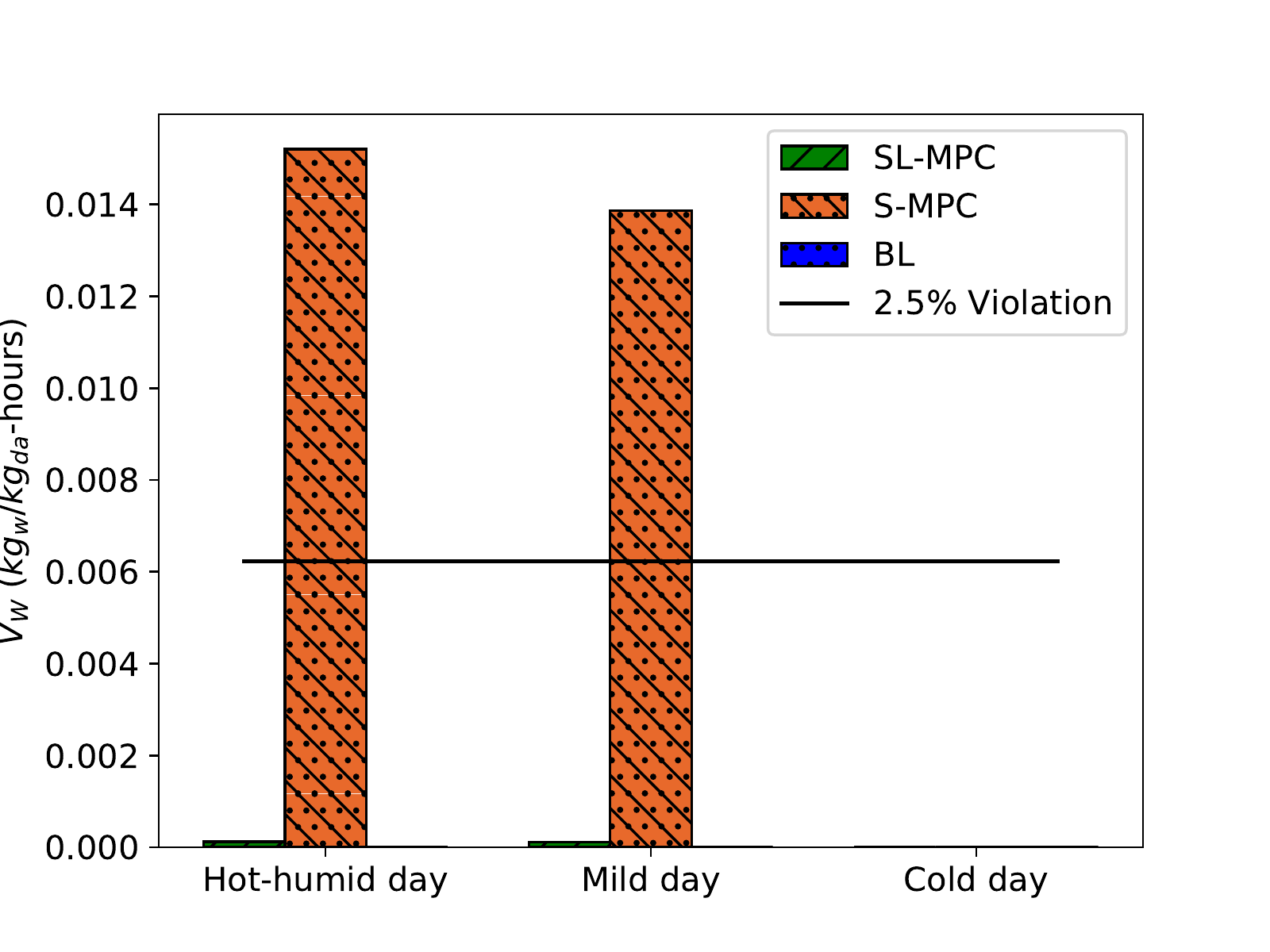}}
	\caption{Comparison of controllers' performance.}
	\label{fig:performance_metrics}
\end{figure}

The performance metrics discussed in Section~\ref{sec:performance_metrics} are computed from simulation data for each of the three controllers, and are shown in Figure~\ref{fig:performance_metrics}. The temperature violation $V_T$ was observed to be minimal for all three controllers, and is therefore not shown in the figure. % with $\base$ having the highest value. This can be attributed to the Proportional Integral (PI) controller in $\base$ that regulates the air flow rate and supply air temperature. Since the temperature violations are minimal the $V_T$ metric is not presented. 
We see from the figure that space humidity is a concern only during hot-humid and mild weather conditions.

Figure~\ref{fig:performance_metrics} shows that $\mpcwh$ outperforms the other two controllers: it consumes the least amount of energy under various outdoor weather conditions with negligible violation in thermal comfort constraints. 

The simulation results discussed in the previous section show that $\mpcwoh$ makes decisions that either leads to thermal comfort violations, or higher energy use when compared to $\mpcwh$, under the following conditions.
\begin{enumerate}
	\item \emph{Mild internal heat load but high outdoor humidity (e.g. spring/summer night)}: In such a condition, $\mpcwoh$ decides that slightly cooler outside air can provide free cooling, but the high humidity of the outside air causes high indoor humidity; see Figures~\ref{fig:humidity_violation}, \ref{fig:comparison_zone_conditions_1_hot}, and \ref{fig:comparison_zone_conditions_1_mild}. Thus, $\mpcwoh$ makes decisions in the interest of reducing energy/cost that leads to violation of humidity constraints. 
	\item \emph{During occupied hours on a mild weather day:} In this case, the $\mpcwoh$ decides to meet the air flow requirement with  a larger fraction of outdoor air due to its small sensible heat, failing to recognize its high latent heat. Because of the high internal sensible load during occupied hours, it uses a low conditioned air setpoint which incidentally reduces humidity of the air supplied, so fortunately no space humidity violations occur. However, the decision is energy inefficient compared to the proposed $\mpcwh$ controller. 
\end{enumerate}

Interestingly, $\mpcwoh$ makes decisions similar to the proposed $\mpcwh$ controller either when (i) the internal heat load is high and outdoor weather is hot and humid, or (ii) the outdoor weather is cold and dry. In the latter case, the cold outdoor is used to provide free (sensible) cooling, and since it is dry there is no risk of space humidity becoming large. In the former scenario, $\mpcwoh$ recognizes that the conditioned air temperature must be low enough to maintain the indoor temperature within allowable limits. That decision has an unintended, but good, side effect of maintaining space humidity even though the controller has no knowledge of humidity.

$\base$ uses conservative set points which ensures that there are no violations in humidity and temperature constraints almost all the time, but leads to higher energy use when compared to the two MPC controllers.
%Conventional rule-based controllers use conservative set points for conditioned air temperature and minimum supply air flow rate, which ensures that humidity violations are not a concern. But with a majority of the current day research focusing towards optimization-based controllers for buildings, it is important to recognize the implications of ignoring humidity and latent heat in the problem formulation, especially in hot and humid climates. Most existing buildings use rule-based controllers and hence do not monitor humidity at various locations in the air loop and in the zones. Incorporating humidity into MPC formulation requires that these sensors be installed and monitored. Thus, it might be a good idea for future building control systems to have these sensors installed so that MPC controllers which incorporate humidity in the formulation can be easily implemented.

These observations provide a basis for a cost-benefit trade-off analysis of the three controllers. One should first note that the proposed $\mpcwh$ requires more sophisticated modeling and additional humidity sensors compared to an MPC controller that ignores humidity/latent heat, and thus is more expensive to use in practice. The additional energy cost savings due to the proposed MPC controller over the naive MPC controller $\mpcwoh$ is small, about 5\%; see Figure~\ref{fig:total_energy_consumption}. The larger difference is space humidity. As discussed previously, in mild outdoor weather conditions such as spring/summer nights, the proposed MPC controller is able to maintain space humidity constraints while the naive MPC controller leads to poor space humidity. Since this large space humidity occurs over many hours in a night (see Figures~\ref{fig:comparison_zone_conditions_1_hot} and \ref{fig:comparison_zone_conditions_1_mild}), and this behavior is likely to repeat every night over the entire season, it may lead to mold growth which can seriously affect occupant health. In fact, additional simulations that are not reported here due to space constraints, show that with larger internal moisture generation, poor space humidity occurs even during occupied hours. Therefore, the benefit of incorporating humidity/latent heat in MPC control of HVAC---or conversely the cost of not doing so---maybe more about occupant health and comfort, and less about energy savings. Additionally, since these benefits (conversely, costs) manifest only over long time periods, experimental evaluations conducted over a short time period, say, less than a few months, may not be adequate to provide a complete assessment.

\section{Conclusion}\label{sec:conclusion}
An MPC controller which incorporates humidity and latent heat in a principled manner is presented. Simulations show that the proposed MPC controller outperforms both a naive MPC controller (that does not consider humidity/latent heat) and a baseline rule-based controller in both energy use and thermal comfort, despite large plant-model mismatch. A thorough comparison for several weather conditions indicate the key advantage of the proposed controller over the naive MPC is not energy savings but humidity control. The naive controller may lead to poor humidity control, especially during mild outdoor weather conditions such as spring/summer nights. Such violations in humidity over long periods of time can cause mold growth and can affect occupant health and comfort.

This study is a first step; there are many avenues for further exploration. A natural extension is to multi-zone buildings. A thorough numerical investigation, for various climate zones and HVAC systems, is also needed. Validating the accuracy of the cooling and dehumidifying coil model against experimental data is yet another avenue for future work.  The underlying optimization problem can also be reformed to guarantee feasibility and convexity. Theoretical properties of the controller need to be investigated as well.

\section*{Acknowledgment}
This research reported here has been partially supported by the NSF through award \# 1646229 (CPS/ECCS) and the State of Florida through a REET (Renewable Energy and Energy Efficiency Technologies) grant.

%\section*{References}
\bibliographystyle{elsarticle-num}
\bibliography{\DiCEbibPATH/Barooah,\DiCEbibPATH/optimization,\DiCEbibPATH/systemid,\DiCEbibPATH/grid,\DiCEbibPATH/building,\DiCEbibPATH/ControlTheory,\NRbibPATH/Raman}

\end{document}